\documentclass[12pt,a4paper]{article}

\usepackage{amsmath}
\usepackage{graphicx,psfrag,epsf,epstopdf}
\usepackage[authoryear,round,semicolon]{natbib}
\usepackage{amssymb}
\usepackage{amsthm}
\usepackage{color}
\usepackage{caption}
\usepackage{booktabs}
\usepackage{bbm}
\usepackage[normalem]{ulem}
\usepackage{hyperref}
\def\conv{\textbf{\emph{conv}}}

\newtheorem{prop}{Proposition}

\newtheorem{algorithm}{Algorithm}

\def\bmu{{\boldsymbol u}}
\def\bmw{{\boldsymbol w}}
\def\bmx{{\boldsymbol x}}
\def\bmy{{\boldsymbol y}}
\def\bmB{{\boldsymbol B}}
\def\bmi{{\boldsymbol i}}

\def\bmd{{\boldsymbol d}}
\def\bmc{{\boldsymbol c}}

\def\IR{{\mathbb{R}}}

\def\u{\textbf{\textit{u}}} \def\x{\textbf{\textit{x}}}  \def\z{\textbf{\textit{z}}}  \def\vv{\textbf{\textit{v}}}
   \def\0{\boldsymbol 0} 

\newcommand\argmax{\operatornamewithlimits{argmax}}

\begin{document}

  \title{\bf Fast computation of Tukey trimmed regions and median in dimension $p>2$}
  \author{Xiaohui Liu\thanks{
  	Xiaohui Liu's research was supported by \textit{NNSF of China (Grant No.11601197, 11461029), China Postdoctoral Science Foundation funded project (2016M600511, 2017T100475), NSF of Jiangxi Province (No.2018ACB21002, 20171ACB21030).}}\hspace{.2cm}\\
    School of Statistics, Jiangxi University of Finance and Economics; \\
    Research Center of Applied Statistics, \\
    Jiangxi University of Finance and Economics\\
    Karl Mosler \\
    Institute of Econometrics and Statistics, University of Cologne\\
    and \\
    Pavlo Mozharovskyi \\
    LTCI, T\'{e}l\'{e}com ParisTech, Universit\'{e} Paris Saclay}

\date{November 8, 2018}

\maketitle

\bigskip
\begin{abstract}
Given data in $\mathbb{R}^{p}$, a Tukey $\kappa$-trimmed region is the set of all points that have at least Tukey depth $\kappa$ w.r.t. the data. As they are visual, affine equivariant and robust, Tukey regions are useful tools in nonparametric multivariate analysis.  While these regions are easily defined and interpreted, their practical use in applications has been impeded so far by the lack of efficient computational procedures in dimension $p > 2$. We construct two novel algorithms to compute a Tukey $\kappa$-trimmed region, a na\"{i}ve one and a more sophisticated one that is much faster than known algorithms. Further, a strict bound on the number of facets of a Tukey region is derived. In a large simulation study the novel fast algorithm is compared with the na\"{i}ve one, which is slower and by construction exact, yielding in every case the same correct results. Finally, the approach is extended to an algorithm that calculates the innermost Tukey region and its barycenter, the Tukey median. Supplementary material is available online.
 \end{abstract}

\vfill

\section{Introduction}

To describe the centrality of a point in a multivariate set of data,  \cite{Tukey75} (see also \cite{Donoho82}) made a celebrated proposal:
Given data ${\cal X} = \{\x_{1},\dots , \x_n\} \subset \mathbb{R}^{p}$ and an additional point $\x$,
\begin{eqnarray}
\label{HDSV}
d(\x) = d(\x | P_{\cal X}) = \inf_{\u \in \mathcal{S}^{p - 1}} \frac{1}{n} \# \left\{i \in \{1,\cdots, n\}: \u^{\top}\x \leq \u^{\top}\x_{i}
\right\}
\end{eqnarray}
measures how central $\x$ is situated w.r.t. the data. Here $\mathcal{S}^{p - 1} = \{\vv\in \mathbb{R}^{p}: \|\vv\| = 1\}$ denotes the unit sphere in $\mathbb{R}^p$, $\#(B)$ the cardinal number of a set $B$,
and $P_{\cal X}$ is the empirical distribution of the data.
As a function of $\x$, \eqref{HDSV} is referred to in the literature as \textit{location depth} or \textit{halfspace depth}, and also, {to recognize the seminal work of \cite{Tukey75}}, as \textit{Tukey depth}. The Tukey depth attains its maximum at the
{\textit{Tukey median set}. For these and the following properties of the Tukey depth, see \textit{e.g.} \cite{ZuoS00a} and \cite{ZuoS00b}. The depth decreases when $\x$ moves on a ray originating from any point of the Tukey median set;}
it vanishes outside the convex hull, $\conv(\x_{1},\dots , \x_n)$, of the data.  Thus, the depth provides a center-outward ordering of points in $\mathbb{R}^{p}$.

The Tukey depth is invariant against a simultaneous affine transformation of $\x$ and the data. Moreover, and most important, its value does not change as long as any of the points $\x, \x_1, \dots, \x_n$ is moved without crossing a hyperplane that is generated by $p$ elements of $\{\x, \x_1, \dots, \x_n\}$. By the last property, the Tukey depth is rather robust against outlying observations; see also \cite{Donoho82}.

A \textit{Tukey $\kappa$-region} $\mathcal{D}(\kappa)$ is the set of points in $\mathbb{R}^{p}$ that have at least Tukey depth $\kappa$,
\begin{equation}\label{TukeyRegion}
  \mathcal{D}(\kappa)=\{\x\in \IR : d(\x) \ge \kappa\}\,,
\end{equation}
$0<\kappa\le 1$. It is a closed convex polyhedron, included in $\conv(\x_{1},\dots , \x_n)$, and hence compact.
Tukey regions are nested, they shrink with increasing $\kappa$.
An empirical distribution is fully characterized by its Tukey regions \citep{StruyfR99}.


In multivariate analysis, a broad nonparametric methodology has been based on Tukey depth and Tukey regions; including multivariate procedures of signs and ranks, order statistics, quantiles, and measures of outlyingness \citep{LiuPS99, Serfling06}. As the Tukey depth is affine invariant and robust, so is any inference based on it.

A Tukey region is a set-valued statistic. Calculating the Tukey $\kappa$-region at some $\kappa$ is an essential part of many statistical tasks, among them
calculating the Tukey median set and its gravity point \citep{Donoho82},
drawing a bagplot and identifying outliers \citep{RousseeuwRT99,HubertRS15},
constructing a multivariate confidence region \citep{YehS97},
measuring multivariate value at risk \citep{CascosM07}, and
solving a risk constraint stochastic program \citep{MoslerB14}.

Moreover, Tukey regions visualize the location, spread and shape of data, particularly in dimension three \citep{LiuPS99}.
This last point will be illustrated at the end of Subsection~\ref{ssec:algBFS}, where we visualize a $3$-dimensional set of real data by a few depth contours.

Many other depth notions have been proposed and used in the literature for descriptive as well as inferential procedures.
Among them the Tukey depth is most important due to its conceptual simplicity, its robustness against outlying data, and its affinity to projection pursuit approaches.

However, most decisive for the possible use of the Tukey depth in applications is, whether the depth and its trimmed regions can be efficiently computed for realistic sample sizes and in higher dimensions.
To calculate the Tukey depth  $d(\x | P_{\cal X})$ of a point $\x$, feasible algorithms have been developed by \cite{RutsR96a} for dimension $p=2$, \cite{RousseeuwS98} for $p=3$, and most recently by \cite{DyckerhoffM16} and \cite{Liu17} for general $p$; see also \cite{MillerRRSSSS03}.

To compute a Tukey $\kappa$-region of given data $\{\x_1,\dots, \x_n\}$ appears to be a much more challenging
task, as it involves a very large number of possible observational hyperplanes to be inspected. By an \textit{observational hyperplane} or \textit{elemental hyperplane} we mean a hyperplane that passes through (at least) $p$ elements of $\{\x_1, \dots, \x_n\}$.
In dimension two the task has been solved by use of a circular sequence, which enumerates all intersections of observational hyperplanes \citep{RutsR96a}.
Throughout the paper we assume that the data is \emph{in general position}, \textit{i.e.}, that any observational hyperplane contains exactly $p$ data points \citep{Donoho82}. Otherwise, the data may be slightly perturbed to meet the assumption.

The Tukey $\kappa$-region $\mathcal{D}(\kappa)$ is the infinite intersection over all directions $\u\in S^{d-1}$ of the inner halfspaces bordered by $\mathcal{H}_{\text{KM}}(\kappa, \u)$. Hereafter,
\begin{eqnarray*}
    \mathcal{H}_{\text{KM}} (\kappa, \u) = \{\z \in \mathbb{R}^{p}: \u^{\top} \z \ge q_{1}(\kappa, \u)\}\,,
\end{eqnarray*}
where $q_{1}(\kappa, \u)$ denotes the sample $\kappa$-quantile of the projections of $\{\x_1, \dots, \x_n\}$ onto $\u$. As a convex polytope, a Tukey region is the intersection of a \textit{finite} number of these halfspaces.
The facets of the polytope lie on the hyperplanes that border the halfspaces. Clearly, by the definition of Tukey depth \eqref{HDSV}, each of these hyperplanes must be an observational hyperplane. Consequently, the Tukey region is completely determined by a finite number of observational hyperplanes. Hence, to calculate it, the key step is to identify those observational hyperplanes that actually include one of the region's facets.
A na\"{i}ve procedure consists in checking all $\binom{n}{p}$ observational hyperplanes.
For a more efficient procedure, we need a strategy to identify those observational hyperplanes that contain the facets.

\cite{HallinPS10} and \cite{PaindaveineS11}, hereafter HPS, point out a direct connection between a Tukey region and multivariate regression quantiles.
Each of these quantiles consists of, in general more than one, parallel hyperplanes, which may contain a facet of the Tukey region.
In their pioneering work, \cite{HallinPS10} show that those directions giving the same set of hyperplanes form a polyhedral cone, and that a finite number of these cones fills $\mathbb{R}^p$.
Each cone is represented by the directions generating its edges, which, again, are finitely many.
HPS propose an algorithm that calculates Tukey regions in dimension $p > 2$ via quantile regression and parametric programming. To guarantee all cones to be addressed, a breadth-first search is used. For details of the implementations, see
\cite{PaindaveineS12a, PaindaveineS12b}. However, these procedures are rather slow.

In the sequel we present two new algorithms, a na\"{i}ve and a more sophisticated one, that calculate Tukey regions in arbitrary dimension $p > 2$. In building the second algorithm, certain combinatorial properties of Tukey regions are derived and exploited that substantially reduce the computational load.
Consequently this algorithm runs much faster and requires much less RAM than the na\"{i}ve algorithm as well as the algorithms by HPS.

Specifically, as a first main result, an upper bound is derived on the number of \textit{non-redundant hyperplanes} of a Tukey region, that is, those observational hyperplanes that contain a facet of the region. The bound is sharp and turns out to be very useful in assessing the computational complexity and performance of the algorithms. To our knowledge of the literature, the bound is new.

The HPS procedures are {slow} for two reasons. First, it appears that both their implementations yield a great number of \emph{redundant directions}, which are normal to an observational hyperplane but provide no facet of the trimmed region. However, all these directions are considered in HPS and used to calculate regions. Given $\kappa$, the HPS procedures \emph{actually calculate $p$ successive regions} instead of one by breadth-first search; but many of these regions have depth $\neq\kappa$ (see \cite{PaindaveineS11}, remark after Theorem 4.2).
Second, the \emph{cone-by-cone} search strategy is both RAM- and time-consuming. This is because, in the HPS procedures, each cone is characterized by its facets and vertices, and facets are identified by $p$-variate vectors. A rather large RAM is required to store these identifiers with sufficient precision.

First, we present a na\"{i}ve combinatorial algorithm (Algorithm~\ref{alg:0}). It serves as a benchmark for our principal fast algorithm (Algorithm~\ref{alg:1}). The na\"{i}ve procedure, in searching for facets' candidates, simply passes through all combinations of $p-1$ observations as the case may be. No memory-consuming structure has to be created, and the computation time is independent of $\kappa$.
In contrast, our fast approach (Algorithm~\ref{alg:1}) uses a breadth-first search {strategy}. However, instead of {covering $\mathbb{R}^p$} \emph{cone-by-cone}, as is done by HPS, it searches the directions \emph{ridge-by-ridge}, where a \emph{ridge} corresponds to a combination of $p-1$ observations in $\mathbb{R}^p$.
This strategy yields only \emph{relevant hyperplanes} (that cut off exactly the required number of observations from  $\mathcal{D}(\kappa)$), and thus examines much fewer cases. Additionally, we store each ridge by the subscripts of its $p-1$ corresponding observations and use some novel tricks to substantially save both RAM and computation time. Obviously, Algorithm~\ref{alg:0} is exact. For Algorithm~\ref{alg:1}, we have no theoretical proof of its exactness though. But we have broad numerical evidence that it computes the exact region. In all our experiments Algorithm~\ref{alg:1} yields precisely the same Tukey region as the exact Algorithm~\ref{alg:0} does.

Similar to HPS, our approach exploits the connection between Tukey trimmed regions and quantile regions pointed out by \cite{KongM12}, \textit{viz.} the fact that the Tukey $\kappa$-region is the intersection of directional $\kappa$-quantile halfspaces, taken over all directions of the unit sphere. Additionally, we derive special combinatorial properties of a Tukey region and make essential use of them in our algorithms.


As they involve only simple operations and no optimization techniques, both
our algorithms are easy to program. They also show high numerical precision even in larger dimension. Particularly, Algorithm~\ref{alg:1}, by its speed and storage efficiency, enables the use of statistical methodology based on Tukey-region statistics.
To investigate the performance of the algorithms, a simulation study (up to dimension $9$) as well as real data calculations are provided. Our procedures have been implemented in C++ and interfaced and visualized in $\texttt{R}$ \citep{Rcore}. They are available in the \texttt{R}-package \texttt{TukeyRegion} \citep{TukeyRegionRpac} and can be downloaded from CRAN.

The \emph{Tukey median} {\citep{Tukey75}} is one of the most famous generalizations of the ordinary median to dimension $p > 1$. It is usually defined \citep{Donoho82} as the average of all points in the Tukey median set. Hence its computation depends essentially on the computation of this innermost Tukey region. As an extension of our approach, we provide an algorithm for fast computation of the Tukey median.

The rest of this paper is organized as follows. Section~\ref{MainIdea} presents an upper bound on the number of non-redundant hyperplanes of a Tukey region, together with some results that are useful for our algorithms. Section~\ref{Algorithm} describes the two novel algorithms, the na\"{i}ve   as well as the fast one.
Section~\ref{sec:comparisons} studies its computational performance, also compared to HPS.
Section~\ref{sec:median} is about computing the Tukey median, and Section~\ref{Conclusions} concludes.

We collect some notions and notations for later reference: Denote $m_{\kappa} = \lceil n \kappa \rceil$ with $\lceil \cdot \rceil$ being the ceiling function.
For any hyperplane $H_{\bmu,\alpha}=\{\bmx\in \IR^p : \bmu'\bmx =\alpha\}$, we say that $H_{\bmu,\alpha}$ \textit{cuts off $m$ observations} if
\begin{equation*}
  \min\bigl\{\#\{i : \bmu'\bmx_i >\alpha\}, \#\{i : \bmu'\bmx_i <\alpha\}\bigr\} = m \,.
\end{equation*}
{If these two numbers are different, the normal pointing to the side with less observations is called the \emph{outer direction} of a hyperplane.} An observational hyperplane that cuts  off exactly $m_{\kappa} - 1$ observations is mentioned as a \emph{relevant hyperplane} of the Tukey $\kappa$-region $\mathcal{D}(\kappa)$; its inner halfspace as a \emph{relevant halfspace}. Obviously,
every non-redundant hyperplane (\textit{i.e.} containing a facet) is also relevant, and
$\mathcal{D}(\kappa)$ is the intersection of all 
relevant halfspaces.

By a $\kappa$-\emph{outside ridge} we mean a $(p-2)$-dimensional affine space
which contains  $p-1$ observations and
is the intersection of two observational hyperplanes, each cutting off, on its lower side, less than $m_\kappa$ observations. Note that these two observational hyperplanes are not unique (see Figure 1 for an illustration in dimension $p=2$) and that only some of the
$\kappa$-outside ridges belong to $\mathcal{D}(\kappa)$.

\section{A bound on the number of facets}
\label{MainIdea}

In this section, as a first principal result, a bound on the number of facets of a Tukey region is derived, that is, on the number of non-redundant halfspaces defining the region.

\begin{figure}[!h]
\captionsetup{width=0.85\textwidth}
\centering
\includegraphics[angle=0,width=5in]{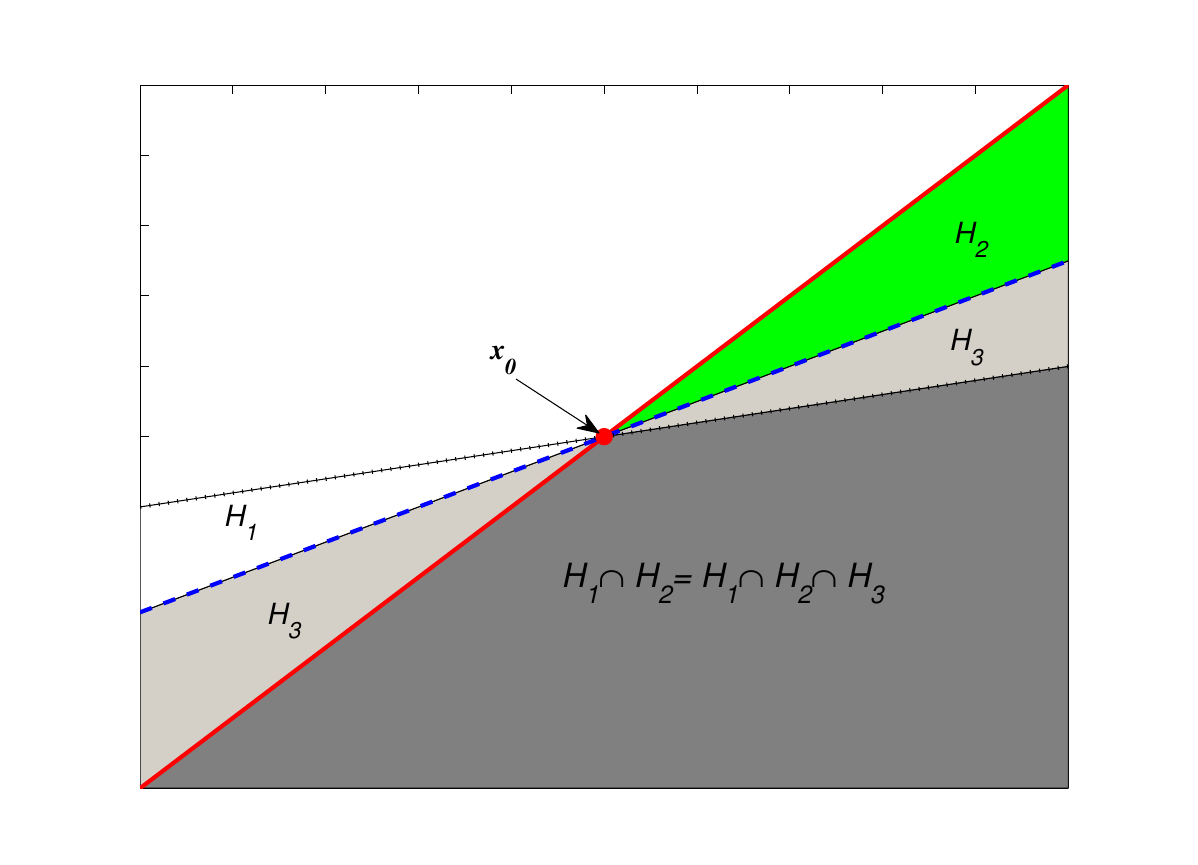}
\caption{Intersection of three relevant halfspaces. Here $\x_0$ denotes the projection of $\x_{i_1}, \x_{i_2},\cdots, \x_{i_{p-1}}$ onto  $\mathbf{V}^\bot$. Clearly, $\bigcap_{k=1}^{3}\mathcal{H}_k = \mathcal{H}_1 \cap \mathcal{H}_2$, and $\mathcal{H}_3$ is redundant.}
\label{fig:proposition1}
\end{figure}

{Assume that the observed data set $\{\x_{1}, \x_{2},\cdots, \x_{n}\} \subset \mathbb{R}^p$, $n> p\ge 2$, is in general position. Consequently, each facet of a Tukey region $\mathcal{D}(\kappa)$ lies on an {observational hyperplane} containing exactly $p$ observations.

\begin{prop}\label{lem:4} Let $\Pi$ be an observational hyperplane.
    If $\Pi$ passes through observations $\x_{i_1}, \x_{i_2},\cdots, \x_{i_p}$ and cuts  off at most $m_{\kappa} - 1$ other observations, then
    $$\Pi \cap \mathcal{D}(\kappa) \subset \emph{\conv}(\x_{i_1}, \x_{i_2},\cdots, \x_{i_p})\,.$$
\end{prop}

{\bf Proof:} Denote $\mathcal{F} = \Pi \cap \mathcal{D}(\kappa)$. If $\Pi$ cuts  off exactly $  m_{\kappa} - 1$ observations and  $\mathcal{F} \neq \emptyset$, then $\mathcal{F}$ must be included in the boundary of $\mathcal{D}(\kappa)$. Now suppose that there exists $\x_0 \in \mathcal{F} \setminus \conv(\x_{i_1}, \x_{i_2},\cdots, \x_{i_p})$. One may use the same technique as \cite{DyckerhoffM16} in moving $\Pi$ around $\x_0$ to exclude all points $\x_{i_j}, j=1,\cdots, p$, from it. This results in depth $d(\x_0) < \kappa$, contradicting with $\x_0 \in \mathcal{D}(\kappa)$.
If $\Pi$ cuts off less than $m_{\kappa} - 1$ observations, then $\mathcal{F} = \emptyset$, and the claim trivially holds.
\hfill $\Box$

Let $\mathbf{V}= {\it span}(\x_{i_1}, \x_{i_2},\cdots, \x_{i_{p-1}})$ be the $(p - 2)$-dimensional vector space spanned by $\{\x_{i_1} - \x_{i_{p-1}}, \x_{i_2} - \x_{i_{p-1}}, \cdots, \x_{i_{p-2}} - \x_{i_{p-1}}\}$, and $\mathbf{V}^\bot$ be its orthogonal complement. (Observe that $\mathbf{V} = \{\0\}$ if $p = 2$.) Consider
the projection of $\mathcal{D}(\kappa)$ onto the two-dimensional vector space $\mathbf{V}^\bot$, which is
$\textit{proj}_{\mathbf{V}^\bot}\bigl(\mathcal{D}(\kappa)\bigr) = \{\x' \in \mathbf{V}^\bot: \x' + \x'' \in \mathcal{D}(\kappa) \text{ with }  \x'' \in \mathbf{V}\}$ . Clearly, this projection is a polygone. Relying on Proposition \ref{lem:4}, one obtains the following result, which  will be useful in constructing our fast algorithm (Subsection~\ref{ssec:algBFS}).

\begin{prop}\label{lem:5}
  Consider the vector space $\mathbf{V}^\bot$ as before, and the $\kappa$-outside ridge containing the observations $\x_{i_1}, \x_{i_2},\cdots, \x_{i_{p-1}}$. The projection of $\x_{i_j}$
   onto $\mathbf{V}^\bot$ is either a vertex or no element of $\textit{proj}_{\mathbf{V}^\bot}\bigl(\mathcal{D}(\kappa)\bigr)$, $j=1,2,\dots, p-1$.
\end{prop}

For $\kappa \in \{1/n, ~2/n, ~\cdots, ~\kappa^*\}$ with $\kappa^* = \sup_{\x \in \mathbb{R}^p} d(\x | P_{\mathcal{X}})$, we see from Proposition \ref{lem:5} and the convexity of $\mathcal{D}(\kappa)$ that there exist at least two relevant hyperplanes each containing the observations $\x_{i_1}, \x_{i_2},\cdots, \x_{i_{p-1}}$ included in this $\kappa$-outside ridge
plus another observation, which is found in the two-dimensional space $\mathbf{V}^\bot$. We will use this fact in Step~2(a) of our Algorithm~1 below.

Moreover, the intersection of the respective halfspaces contains the corresponding $(p-2)$-dimensional affine space.
It is easy to see that among the halfspaces bordered by them two exist, say $\mathcal{H}_1$ and $\mathcal{H}_2$, such that $\mathcal{D}(\kappa) \subset \mathcal{H}_1 \cap \mathcal{H}_2$. Figure~\ref{fig:proposition1} illustrates this; it also demonstrates that a relevant hyperplane can be redundant. Hence, we have the following result.

\begin{prop}\label{lem:6}
Consider $\kappa \in \{1/n,~ 2/n,~\cdots, ~\kappa^*\}$.
If the Tukey region $\mathcal{D}(\kappa)$ is \emph{not} a singleton, the number of its facets (= number of its non-redundant hyperplanes) is bounded from above by $2{n \choose p - 1}/p$.
\end{prop}

{\bf Proof:} As discussed above, if  $\mathcal{D}(\kappa)$ is not a singleton, each $\kappa$-outside ridge can participate in the construction of no more than two facets. Combinatorially, the number of such $\kappa$-outside ridges is bounded from above by ${n \choose p - 1}$. On the other hand, each $(p - 1)$-dimensional facet lies on an observational hyperplane which corresponds to exactly $p$ $\kappa$-outside ridges. From this follows that at most $2{n \choose p - 1}/p$ facets exist. \hfill $\Box$}

By the convexity of the Tukey region, an outer direction vector yields at most one of its facets.
In this sense, Proposition~\ref{lem:6} actually also provides an upper bound for the number of outer directions of facets. It is useful in assessing the performance of an algorithm, and will be used in Step~4 of Algorithm~\ref{alg:1}

It is worth mentioning that the upper bound $2{n \choose p - 1}/p$ is \emph{attainable} and thus {cannot be further improved}. For instance, let $n=p+1$. Then the Tukey region at depth $\kappa=\frac{1}{p + 1}$ is the convex hull of the data. It has $p + 1$ facets, which equals the upper bound, $2{p + 1 \choose p - 1}/p = p + 1$. Though the bound is sharp, in many instances the number of facets comes out to be much smaller, as we will see from the numerical study below.


Propositions~\ref{lem:5} and~\ref{lem:6} reveal further important properties of $\kappa$-outside ridges, which are useful in constructing a fast algorithm to calculate $\mathcal{D}(\kappa)$ for $\kappa \in \{1/n, ~2/n,~\cdots, ~\kappa^*\}$:
\begin{itemize}
  \item Every $\kappa$-{outside ridge} can be utilized to compute at least two directions that are normal to relevant hyperplanes.
  \item {The $\kappa$-outside ridges are connected with each other in the following sense: Given a $\kappa$-{outside ridge}, we may consider its defining observations $\x_{i_1}, \x_{i_2},\cdots, \x_{i_{p-1}}$ and add another observation $\x^*$, so that the hyperplane through $\x_{i_1}, \x_{i_2},\cdots, \x_{i_{p-1}}, \x^*$ is relevant.
      Then another $\kappa$-{outside ridge} is obtained by replacing one of the defining observations with $\x^*$. This can be utilized in the computation of $\mathcal{D}(\kappa)$.}
  \item Observe that a $\kappa$-{outside ridge} is also a $\kappa'$-\emph{outside} for all $\kappa' \in \{\frac{  m_{\kappa} + 1}{n}, ~\frac{  m_{\kappa} + 2}{n}, ~\cdots, ~\kappa^*\}$. This is for example true for the observations projected into the darkest area of Figure~\ref{fig:proposition1}. Each of them can be used in the construction of a $\kappa'$-outside ridge.
      Hence, if one has to calculate {more than one region},
        one may store all $\kappa$-{outside ridges} during the computation of $\mathcal{D}(\kappa)$ and recycle them when computing a more central trimmed region $\mathcal{D}(\kappa')$ with $\kappa' > \kappa$.
\end{itemize}

\section{Algorithms}\label{Algorithm}

This section presents two new algorithms to compute a Tukey region of given depth.
We start by introducing Algorithm~\ref{alg:0} in Subsection~\ref{ssec:algComb}, which is a na\"{i}ve application of Proposition~\ref{lem:5}. Algorithm~\ref{alg:0} is simple and intuitive; it will later serve to verify the correctness of Algorithm~\ref{alg:1}. After this, in Subsection~\ref{ssec:algBFS}, we describe the fast algorithm, Algorithm~\ref{alg:1}.

\subsection{The na\"{i}ve combinatorial algorithm}\label{ssec:algComb}

Algorithm~\ref{alg:0} simply passes through all combinations $\{\x_{i_1},\cdots,\x_{i_{p-1}}\}$ of $p-1$ out of $n$ points and searches for hyperplanes passing through $p$ points and cutting  off exactly $m_\kappa - 1$ observations. To do this, first for each choice of observations $\x_{i_1},\cdots,\x_{i_{p-1}}$, the $(p-2)$-dimensional  vector space spanned by them is calculated (Step~2a), and the sample is projected onto its orthogonal complement, which is a two-dimensional vector space (Step~2b). Then the search narrows down to finding two lines in this plane passing through the point to which
$\x_{i_1},\cdots,\x_{i_{p-1}}$ are projected and another point from the sample and cutting  off $m_\kappa - 1$ observations (Step~2c). Figure~\ref{fig:algBfs2} (right) visualizes the set of hyperplanes found during one iteration in Steps~2a to 2c.
Each found hyperplane is stored
as a number to the basis $n$. This requires $\lceil p\log_2(n)\rceil$ binary digits (bits) of memory space (Step~2d).
Note that, under the assumption of general position made above, ties cannot occur, as any hyperplane contains at most $p$ observations.
 Since the number of those combinations is ${n \choose p - 1}$ and search in each of them is algorithmically dominated by the angle-sorting procedure having time complexity $O\bigl(n\log (n)\bigr)$, the time complexity of the algorithm amounts to $O\bigl(n^{p}\log (n)\bigr)$. Clearly, this presumes that the time complexity of Step~2d(iii) is not larger than that of Step~2d(i), which can be achieved by using appropriate store-search structures such as a search tree (access time complexity $O\bigl(p\log(n)\bigr)$) or a binary hypermatrix (access time complexity $O(1)$). As Algorithm~\ref{alg:0} does not take account of any space ordering it requires very little memory and saves computation time, which otherwise is needed for multiple access of search structures and may grow substantially with $n$ and $p$. In addition, due to the same reason its execution time only negligibly depends on the geometry of the data cloud and the depth value $\kappa$.

\setcounter{algorithm}{-1}

\begin{algorithm}[Na\"{i}ve combinatorial algorithm] \label{alg:0} \quad
\begin{enumerate}
    \item[] {\bf Input:} ${\bf{\textit{x}}}_1, \cdots ,\x_n\in\mathbb{R}^p$, $\kappa$.
    \item[] {\bf Step 1.} Set ${\mathcal H}_{\kappa}=\emptyset$.
    \item[] {\bf Step 2.} For each subset $\{i_1,...,i_{p-1}\}=I\subset\{1,\cdots,n\}$ do:
    \begin{enumerate}
    	\item Consider the plane normal to $\text{span}(\bmx_{i_1},...,\bmx_{i_{p-1}})$ and find a basis of it. Let $\bmB_I$ denote the \emph{basis matrix}, that is the $(p\times 2)$ matrix $\bmB_I$ containing the two basis vectors as columns.
    	\item Compute $\bmy_i=\bmB_I^{\top}\bmx_i$ for $i=1,...,n$.
    	\item Find a subset $I_\kappa \subset \{1,\cdots,n\}\setminus I$ such that for each $i_0\in I_\kappa$ holds $\#\{j\,:\,\bmu^{\top}\bmy_j > \bmu^{\top}\bmy_{i_0}\,,\,j=1,...,n\} = m_{\kappa} - 1$ whenever $\bmu^{\top}(\bmy_{i_1}-\bmy_{i_0})=0$.
    	\item For each $i_0\in I_\kappa$ do:
    	\begin{enumerate}
    		\item $(k_1,...,k_p)=\text{sort}\bigl(i_0,i_1,...,i_{p-1})$.
           \item Compute $h=k_1 + k_2 n + k_3 n^2 + ... + k_p n^{p-1}$.
           \item If $h\notin\mathcal{H}_{\kappa}$ then add $h$ to $\mathcal{H}_{\kappa}$.
    	\end{enumerate}
    \end{enumerate}
    \item[] {\bf Output:} ${\mathcal H}_{\kappa}$.
\end{enumerate}
\end{algorithm}


\subsection{The ridge-by-ridge breadth-first search strategy}\label{ssec:algBFS}

In this subsection we present our main procedure, Algorithm~\ref{alg:1}, that computes the Tukey $\kappa$-region for given $\kappa$ in a fast way. The algorithm has been implemented in function \texttt{TukeyRegion} of the \texttt{R}-package \texttt{TukeyRegion} and can be downloaded from CRAN. Its output depends on the user's request.
A detailed description of the algorithm with reference to user-accessible options is given below.

To \emph{find all relevant hyperplanes} of a Tukey $\kappa$-region, Algorithm~\ref{alg:1} follows the breadth-first search idea. This idea can be briefly explained as follows: push an initial set of ridges into the queue, retrieve a ridge from the queue and push all those
{``neighbors''} into the queue that have not been seen before (which will be specified below in Step 4); continue until the queue is empty (Steps~1 to 5). After this has been done, the region is constructed as an intersection of {relevant} halfspaces. For this, firstly, an inner point of the region has to be found, which is used for a dual transformation of the {relevant} hyperplanes (Step~6); then the QHULL~\citep{BarberDH96} algorithm is run to \emph{eliminate the redundant hyperplanes} (Step~7). Vertices, facets and the barycenter of the
Tukey region are computed from the remaining {(non-redundant)} hyperplanes (Step~8). The input consists of the observations, the depth level $\kappa$ and a precision parameter $\epsilon$.

\begin{algorithm}[Algorithm for computing the Tukey region] \label{alg:1} \quad
\begin{enumerate}
    \item[] {\bf Input:} ${\bf{\textit{x}}}_1, \cdots ,\x_n\subset\mathbb{R}^p$, $\kappa$, $\epsilon$.
    \item[] {\bf Step 1.} {\bf Initialization:}
    \begin{enumerate}
    	\item[] Set ${\mathcal A}=(\texttt{false}_n)^{p-1}$, ${\mathcal H}_{\kappa}=\emptyset$, an empty queue ${\mathcal Q}$.
    \end{enumerate}
    \item[] {\bf Step 2.} {\bf Construct initial set of ridges:}
    \begin{enumerate}
    	\item Find a subset $\{i_1,...,i_{p-1}\}=I\subset\{1,\cdots,n\}$ such that $(\bmx_{i_1},...,\bmx_{i_{p-1}})$ define a ridge of $\text{conv}(\bmx_1,...,\bmx_n)$. Set $A\bigl(\text{sort}(I)\bigr)=\texttt{true}$ and push $\text{sort}(I)$ into $\mathcal{Q}$.
    	\item Compute a basis matrix $\bmB_I$ of the plane normal to $\text{span}(\bmx_{i_1},...,\bmx_{i_{p-1}})$.
    	\item Compute $\bmy_i=\bmB_I^{\top}\bmx_i$ for $i=1,...,n$.
    	\item Find two indices $l_1$, $l_2$ such that holds $\#\{j\,:\,\bmu_k^{\top}\bmy_j > \bmu_k^{\top}\bmy_{l_k}\,,\,j=1,...,n\} = m_\kappa - 1$ whenever $\bmu_k^{\top}(\bmy_{l_k}-\bmy_{i_1})=0$ for $k=1,2$.

    For $k=1,2$ and for each subset $J\subset I$ of order $p-2$, set $A\bigl(\text{sort}(J\cup \{l_k\})\bigr)=\texttt{true}$ and push $\text{sort}(J\cup \{l_k\})$ into $\mathcal{Q}$.
    	\item For $k=1,2$ and for each $l$ such that holds $\bmu_k^{\top}\bmy_l > \bmu_k^{\top}\bmy_{l_k}$ and for each subset $J\subset I$ of order $p-2$, set $A\bigl(\text{sort}(J\cup \{l\})\bigr)=\texttt{true}$ and push $\text{sort}(J\cup \{l\})$ into $\mathcal{Q}$.
    \end{enumerate}
    \item[] {\bf Step 3.} Retrieve a ridge $I=\{i_1,\cdots,i_{p-1}\}$ from ${\mathcal Q}$.
    \item[] {\bf Step 4.} {\bf Spread to neighboring ridges:}
    \begin{enumerate}
    	\item Compute a basis matrix $\bmB_I$ of the plane normal to $\text{span}(\bmx_{i_1},...,\bmx_{i_{p-1}})$.
    	\item Compute $\bmy_i=\bmB_I^{\top}\bmx_i$ for $i=1,...,n$.
    	\item Find a subset $I_{\kappa}\subset\{1,\cdots,n\}\setminus I$ such that for each $i_0\in I_{\kappa}$ holds $\#\{j\,:\,\bmu^{\top}\bmy_j > \bmu^{\top}\bmy_{i_0}\,,\,j=1,...,n\} = m_\kappa - 1$ whenever $\bmu^{\top}(\bmy_{i_1}-\bmy_{i_0})=0$.
    	\item For each $i_0\in I_{\kappa}$ do:
    	\begin{enumerate}
    		\item If $(I\cup \{i_0\})\notin\mathcal{H}_\kappa$ then add $(I\cup \{i_0\})$ to $\mathcal{H}_\kappa$.
           \item For each subset $J\subset I_{\kappa}$ of order $p-2$ do:
           	\begin{enumerate}
           		\item[] If $A\bigl(\text{sort}(J\cup \{i_0\})\bigr)=\texttt{false}$ then set $A\bigl(\text{sort}(J\cup \{i_0\})\bigr)=\texttt{true}$ and push $\text{sort}(J\cup \{i_0\})$ into $\mathcal{Q}$.
           	\end{enumerate}
    	\end{enumerate}
    \end{enumerate}
    \item[] {\bf Step 5.} If ${\mathcal Q}$ is not empty then go to {\bf Step~3}, else go to the following step.
    \item[] (So far, all $p$-tuples of observations that define \emph{relevant halfspaces}
    are stored in $\mathcal{H}_\kappa$. The intersection of these halfspaces is the Tukey $\kappa$-region. We give details below.)
    \item[] {\bf Step 6.} {\bf Find an inner point of the region:}
    \begin{enumerate}
    	\item For each $(i_1,...,i_p)\in\mathcal{H}_\kappa$ ($l=1,...,n_\kappa=\#\mathcal{H}_\kappa$) do:
    	\begin{enumerate}
    		\item Find $\bmu_l\perp\text{span}(\bmx_{i_1},...,\bmx_{i_p})$ such that $\#\{j\,:\,\bmu_l^{\top}\bmx_j > \bmu_l^{\top}\bmx_{i_1}\,,\,j=1,...,n\} = m_\kappa - 1$ and $\|\bmu_l\|_2=1$.
    		\item Compute $b_l=\bmu_l^{\top}\bmx_{i_1}$.
    	\end{enumerate}
    	\item Compute $\bmx_0=\argmax_{\bmx\in\mathbb{R}^p}\{\bmx^{\top}(1,0,...,0)^{\top}\,:\,\bmu_l^{\top}\bmx\le b_l - \epsilon\,,\,l=1,...,n_\kappa\}$.
    	\item If $\bmx_0$ cannot be found then {\bf stop}. 
    \end{enumerate}
    \item[] {\bf Step 7.} {\bf Eliminate redundant halfspaces:}
    \begin{enumerate}
    	\item For $j=1,...,n_\kappa$ do:
    	\begin{enumerate}
    		\item[] $\bmw_j= \frac 1{b_j - \u_j^{\top}\x_0} \u_j$.
    	\end{enumerate}
    	\item $(\bmi_1,...,\bmi_{n^v_\kappa})^{\top} = \text{QHULL}(\bmw_1,...,\bmw_{n_\kappa})$ with $\bmi_j=(i_{j1},...,i_{jp})$.
    \end{enumerate}
    \item[] {\bf Step 8.} {\bf Compute elements that define the region:}
    \begin{enumerate}
    	\item For $j = 1,...,n^v_\kappa$ do:
    	\begin{enumerate}
    		\item[] $\vv_j = \bigl((\bmu_{i_{j1}},...,\bmu_{i_{jp}})^{\top}\bigr)^{-1} (b_{i_{j1}},...,b_{i_{jp}})^{\top}$.
    	\end{enumerate}
    	\item For each $i\in\text{unique}(\bmi_1,...,\bmi_{n^v_\kappa})$ ($j=1,...,n^f_\kappa$) do:
        	\begin{enumerate}
    		\item $\bmd_j = \bmu_i$.
    		\item $t_j = b_i$.
    	\end{enumerate}
    	\item $\bmc=\text{ave}\bigl(\text{conv}(\vv_1,...,\vv_{n^v_\kappa})\bigr)$, the barycenter of $\text{conv}(\vv_1,...,\vv_{n^v_\kappa})$.
    \end{enumerate}
    \item[] {\bf Output:}
    \begin{enumerate}
    	\item Vertices: $\mathcal{V} = \{\vv_1,...,\vv_{n^v_\kappa}\}$.
    	\item Facets' (non-redundant) hyperplanes: $\mathcal{F} = \{\bmd_1,...,\bmd_{n^f_\kappa}\}$ (outside-pointing normals) and $\mathcal{T}=\{t_1,...,t_{n^f_\kappa}\}$ (thresholds on these normals).
    	\item Barycenter: $\bmc$.
    \end{enumerate}
\end{enumerate}
\end{algorithm}

In {\bf Step~1}, $(\texttt{false}_n)^{p-1}$ is a $(p-1)$-dimensional logical matrix having format $n \times \cdots \times n$, that is,
in the beginning an $n$-dimensional vector of logical zeros, brought to power $p-1$ as a Cartesian product.
(\textit{E.g.}, if $p = 4$ this is a cube of size $n \times n \times n$.)
Indeed, only one upper corner of this matrix is used, which includes those cells having strictly decreasing subscripts. Further, a single bit of RAM suffices to store a logical value, which amounts to eight values per byte. However, this matrix is memory demanding when $n$ and $p$ are large. Fortunately, in this last case the matrix is very probable to be sparse, and thus some dynamic storing structure may be used, \textit{e.g.} a search tree. $\mathcal{H_\kappa}$ is a set for storing {relevant} hyperplanes as $p$-tuples of integer numbers, and $\mathcal{Q}$ is literally a queue {of ridges} supporting operations of pushing an element on one side and retrieving it from another one.

\begin{figure}[!t]
\begin{center}
	\includegraphics[scale=1.15]{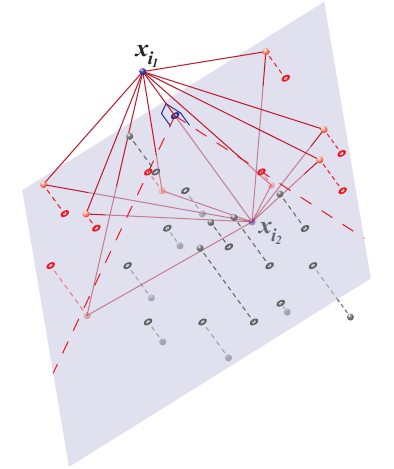} \includegraphics[scale=1.15]{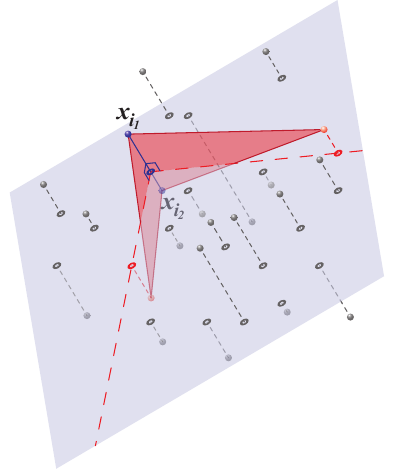}
	\caption{Left: Step~2 of Algorithm~\ref{alg:1}, solid red lines indicate initial ridges. \\ Right: The generic Step~2 (a)~--~(c) of Algorithm~\ref{alg:0} and Step~4 (a)~--~(c) of Algorithm~\ref{alg:1}, solid red lines indicate new ridges added to the queue.}
	\label{fig:algBfs2}
\end{center}
\end{figure}

{\bf Step~2} aims at finding a set of $(p-1)$-tuples defining ridges to be used as starting points for the algorithm. First, a ridge of the convex hull of $\{\bmx_1,...,\bmx_n\}$ defined by, say, $\{\x_{j_1}, \x_{j_2}, \cdots, \x_{j_{p-1}}\}$ is found (Step~2a). This is implemented in the QHULL algorithm. Further, using the logic of Steps~2a to 2c of Algorithm~\ref{alg:0}, two points $\bmx_{l_1}$ and $\bmx_{l_2}$ are found (Steps~2b to 2d). Thus, each of the two hyperplanes defined by $\{\x_{j_1}, \x_{j_2}, \cdots, \x_{j_{p-1}},\bmx_{l1}\}$ and $\{\x_{j_1}, \x_{j_2}, \cdots, \x_{j_{p-1}},\bmx_{l2}\}$ is relevant.
This is always guaranteed as long as the sample is in general position and $m_\kappa\le \lfloor\frac{n - (p - 1)}{2}\rfloor$; here $\lfloor\cdot \rfloor$ denotes the floor function. Then, all $(p - 1)$-tuples containing $(p - 2)$ points out of $\{\x_{j_1}, \x_{j_2}, \cdots, \x_{j_{p-1}}\}$ and one of the points cut off by or lying in one of these hyperplanes are chosen. Together with the ridge on the convex hull of the data set, this gives exactly $1 + 2m_\kappa(p - 1)$ initial ridges, {see Figure~\ref{fig:algBfs2} (left) for illustration}.


{\bf Steps~3} and {\bf 5} wrap the search step procedure of Step~4 by implementing the queue. Step~3 retrieves a ridge to be processed from the head of the queue while Step~5 returns to Step~3 if the queue contains at least one element.

{\bf Step~4} provides the identification of neighboring ridges and, by that, controls the ridge-by-ridge search strategy. First, a set $I_{\kappa}$ of indices is found that defines, together with the current ridge $\{\x_{j_1}, \x_{j_2}, \cdots, \x_{j_{p-1}}\}$, a relevant hyperplane.
(Here we follow the logic of Steps~2a to 2c of Algorithm~\ref{alg:0} as before.)
The information gained by each $i_0\in I_{\kappa}$ is twofold. First, an $i_0$ defines a {relevant} hyperplane determined by a $p$-tuple $\{\x_{i_1}, \x_{i_2}, \cdots, \x_{i_{p-1}},\bmx_{i_0}\}$. We check whether this hyperplane is visited for the first time and add it to $\mathcal{H}_\kappa$ if this is the case (Step~4d(i)). Second, each such hyperplane contains $p$ ridges defined by $(p-1)$-tuples of points. $p-1$ of these ridges -- those containing $i_0$ -- can potentially lead to a hyperplane not visited before. Thus we check each of these $(p-1)$ ridges whether it has been visited before or not. If a ridge has not been visited we add it to the queue $\mathcal{Q}$ and mark it as visited in $\mathcal{A}$ (Step~4d(ii)). The set of ridges found in Step~4 of a single iteration is visualized in Figure~\ref{fig:algBfs2} (right).

As the total number of {relevant} hyperplanes (after the first five steps of Algorithm~\ref{alg:1} have been performed) can potentially be as large as their maximum number, the complexity of the algorithm is the same as that of Algorithm~\ref{alg:0}, \textit{i.e.} $O\bigl(n^p\log(n)\bigr)$. On the other hand, this worst case happens only for degenerate data sets. Thus it is reasonable to expect that in most cases the number of {relevant hyperplanes} is substantially lower than the upper bound, and this number actually defines the computation speed of the algorithm. Some empirical insights to this question will follow in Section~\ref{sec:comparisons}.

{\bf Steps~1} to~{\bf 5} aim to compute {relevant} hyperplanes. We will use this set to check the correctness of the algorithm when comparing it empirically with the output of Algorithm~\ref{alg:0} in Subsection~\ref{ssec:validation}. The implementation of Algorithm~\ref{alg:1} in the \texttt{R}-package \texttt{TukeyRegion} performs these five first steps always independently of the chosen options. The following steps are optional. If the algorithm is terminated after Step~5, only the {relevant} hyperplanes are output, each as a $p$-tuple $\{{j_1}, {j_2}, \cdots, {j_{p}}\}$ of indices of points from the sample.



{\bf Step~6} searches for a point strictly belonging to the interior of the Tukey region. If such a point cannot be found then
the algorithm stops. (See also Section~\ref{sec:median}.)
Step~6 is performed either if the interior point is explicitly requested (flag \texttt{retInnerPoint} is set) or if any option for the following steps is chosen.

Each element of $\mathcal{H}_\kappa$ is a $p$-tuple of indices $({j_1}, {j_2}, \cdots, {j_{p}})$ defining a halfspace by the hyperplane containing $\{\bmx_{j_1}, \bmx_{j_2}, \cdots, \bmx_{j_{p}}\}$. First, for each of these hyperplanes the normal pointing outside the Tukey region and the threshold defining the hyperplane's position are calculated (Step~6a); each of them defines a condition for the interior point. Then, the inner point $\bmx_0$ of the region is searched by means of linear programming \citep[\texttt{R}-package \texttt{Rglpk} used here;][]{RglpkRpac} as a point satisfying these conditions (Step~4b).
If the inner point is not found this means that the Tukey region of depth $\kappa$ has (numerically, with \emph{precision} $\epsilon$) zero volume or does not exist. In this case the algorithm stops.


{\bf Step~7} {determines non-redundant} hyperplanes. It is performed either if these are requested (flag \texttt{retHyperplanesNR}) or if any option for the following step is chosen. First, a duality transformation is applied. It represents each halfspace by a vector  $\bmw_j$, which is its outer normal multiplied by the inverse distance of the hyperplane to the inner point (Step~7a). Second, the QHULL algorithm is applied to the set of all $\bmw_j$'s. It returns the convex hull of the $\bmw_j$'s, an $n^v_\kappa\times p$ matrix, where each row defines a facet by indices of $p$ points (Step~7b). Since the data is in general position, each facet of this convex hull is defined by exactly $p$ points.

{\bf Step~8} computes the elements of the region; it is the step that provides the practically important output. This step depends on whether region's vertices (flag \texttt{retVertices} is set), facets (flag \texttt{retFacets} is set) and/or barycenter (flag \texttt{retBarycenter} is set) are requested and requires the successful execution of all preceding steps.
 Each facet of the convex hull of $\bmw_j$'s in the dual space (corresponding to a row of the matrix returned by QHULL in Step~7b) defines a vertex of the Tukey region, \textit{i.e.} each such vertex is computed as an intersection of $p$ non-redundant hyperplanes (Step~8a).

 The facets of the Tukey region are contained in {non-redundant} hyperplanes defined by pairs $(\bmd_j,t_j),\,j=1,...,n^f_\kappa$ obtained as
  non-repeating entries of the matrix $(\bmi_1,...,\bmi_{n^v_\kappa})^{\top}$ (Step~8b).
  To obtain facets as polygons one can apply QHULL to the set of regions' vertices $\{\vv_1,...,\vv_{n^v_\kappa}\}$. In addition, the barycenter of a Tukey region can be computed as the weighted average of the triangulated (doable by QHULL as well) region, where points are the means of vertices of simplices ($p$-dimensional triangles) and weights are the volumes of these simplices (Step~8c).


\begin{figure}[!ht]
	\begin{center}
\def\ww{.425\textwidth}
\def\www{.475\textwidth}
\fbox{\vbox{\hsize=\www\includegraphics[keepaspectratio=true,width = \ww, trim = 1mm 1mm 1mm 1mm, clip,scale=0.75, page = 1]{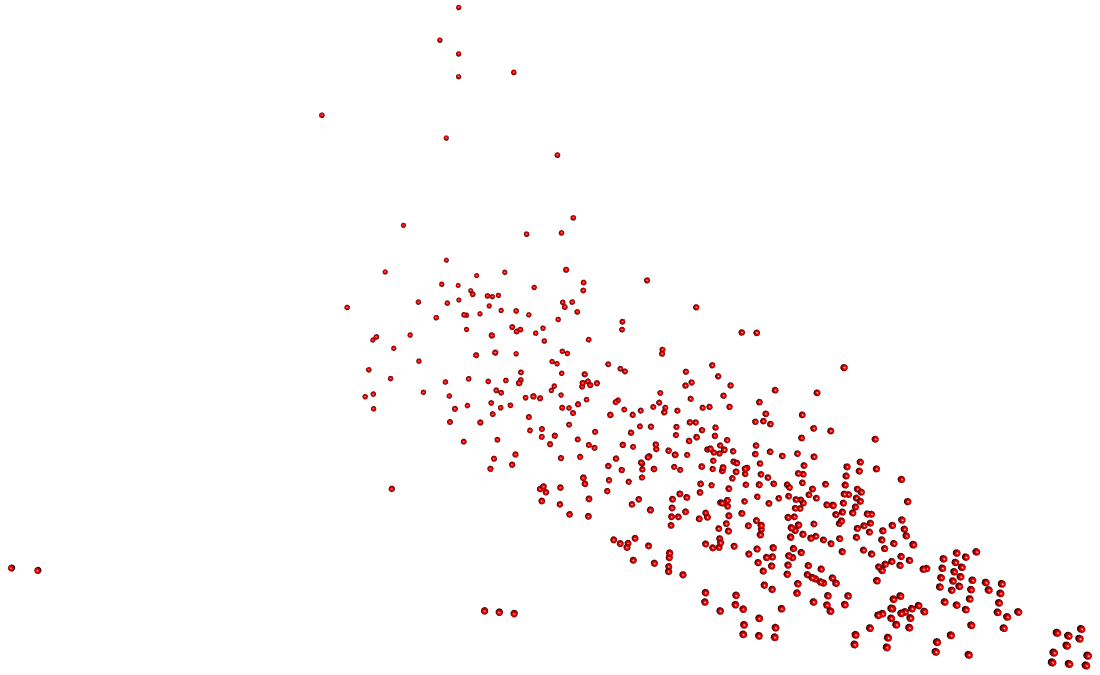}\\
{\small The data \phantom{g} }}}
\fbox{\vbox{\hsize=\www\includegraphics[keepaspectratio=true,width = \ww, trim = 1mm 1mm 1mm 1mm, clip,scale=0.75, page = 1]{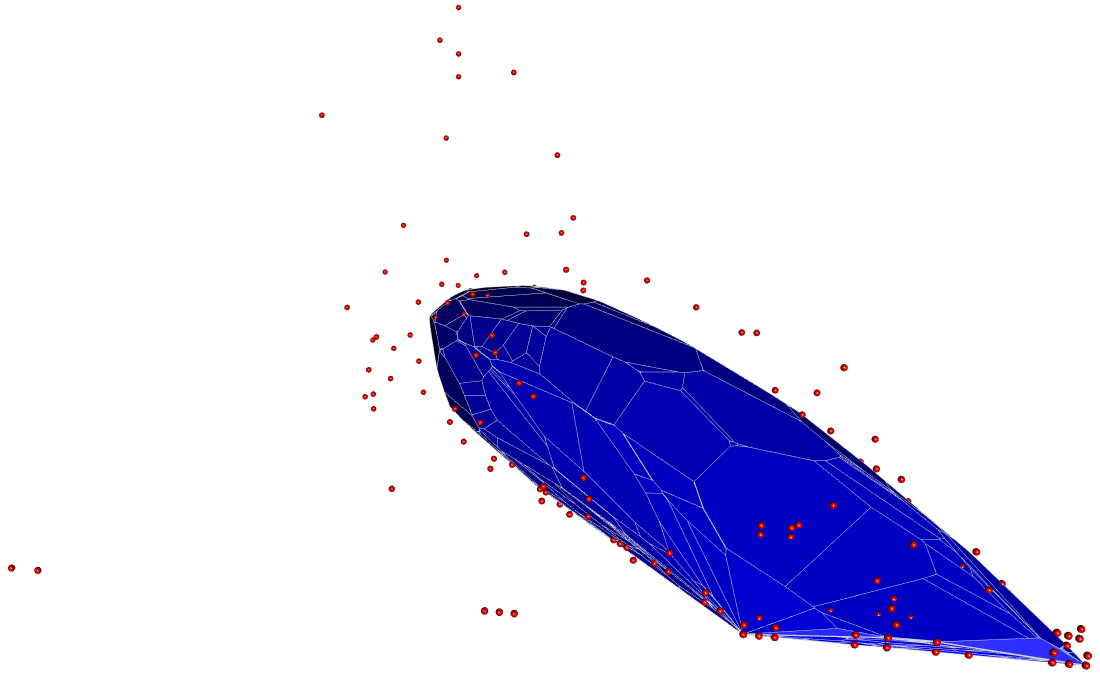}\\
{\small Tukey region for $\kappa = 0.025$ }}}

\fbox{\vbox{\hsize=\www\includegraphics[keepaspectratio=true,width = \ww, trim = 1mm 1mm 1mm 1mm, clip,scale=0.75, page = 1]{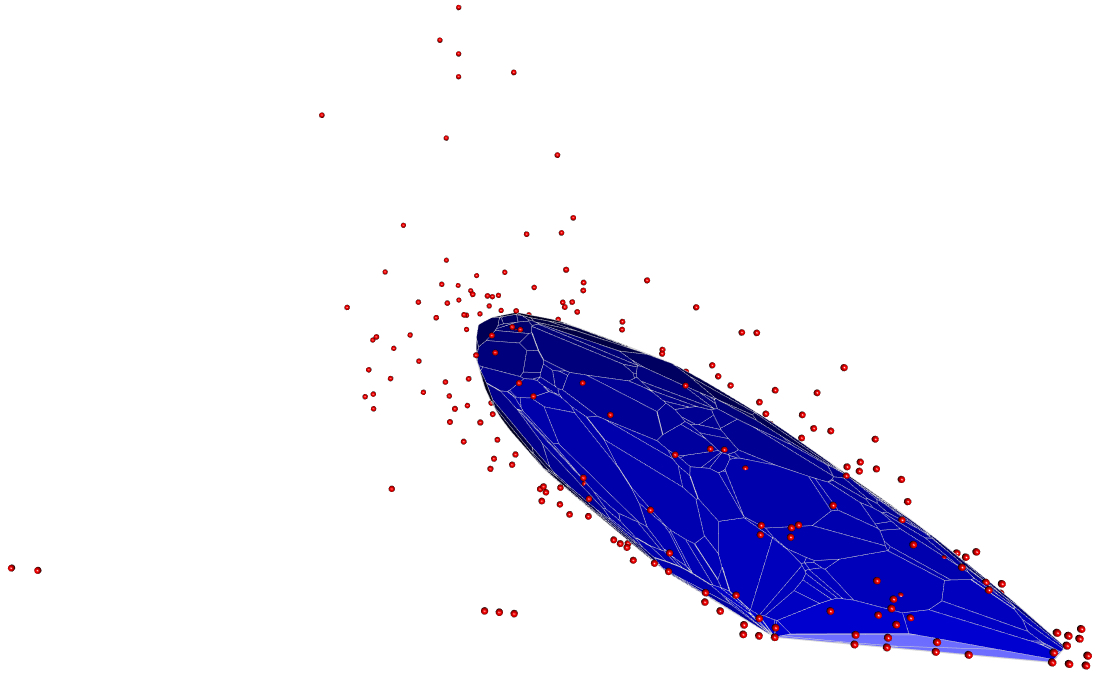}\\
{\small Tukey region for $\kappa = 0.05$ }}}
\fbox{\vbox{\hsize=\www\includegraphics[keepaspectratio=true,width = \ww, trim = 1mm 1mm 1mm 1mm, clip,scale=0.75, page = 1]{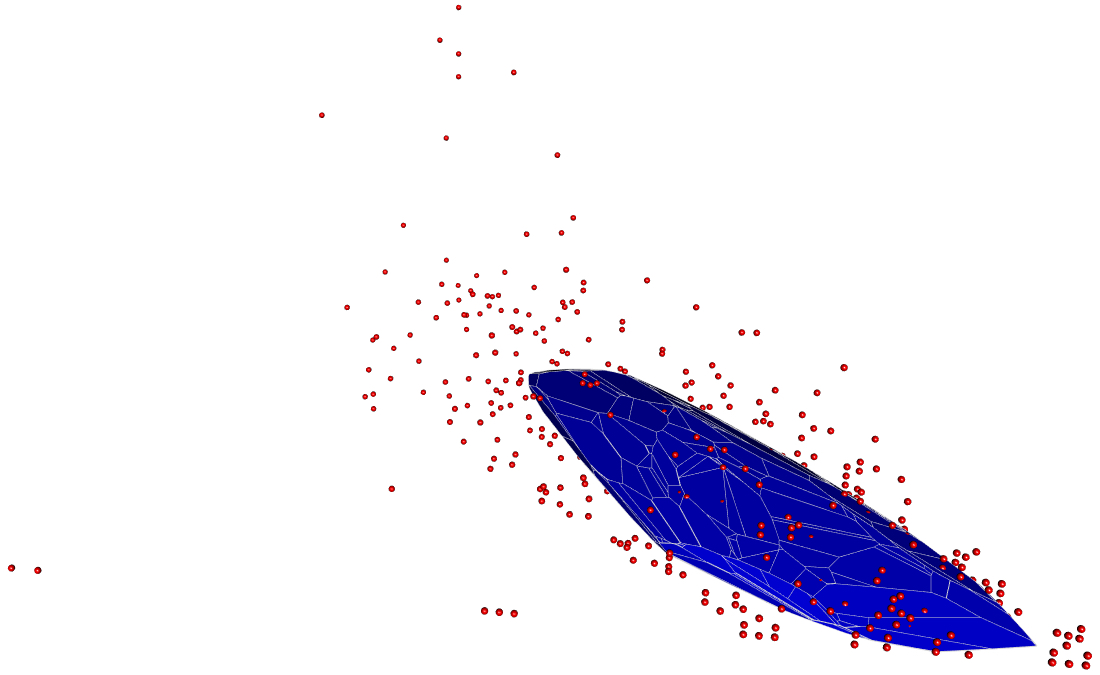}\\
{\small Tukey region for $\kappa = 0.1$ }}}

\fbox{\vbox{\hsize=\www\includegraphics[keepaspectratio=true,width = \ww, trim = 1mm 1mm 1mm 1mm, clip,scale=0.75, page = 1]{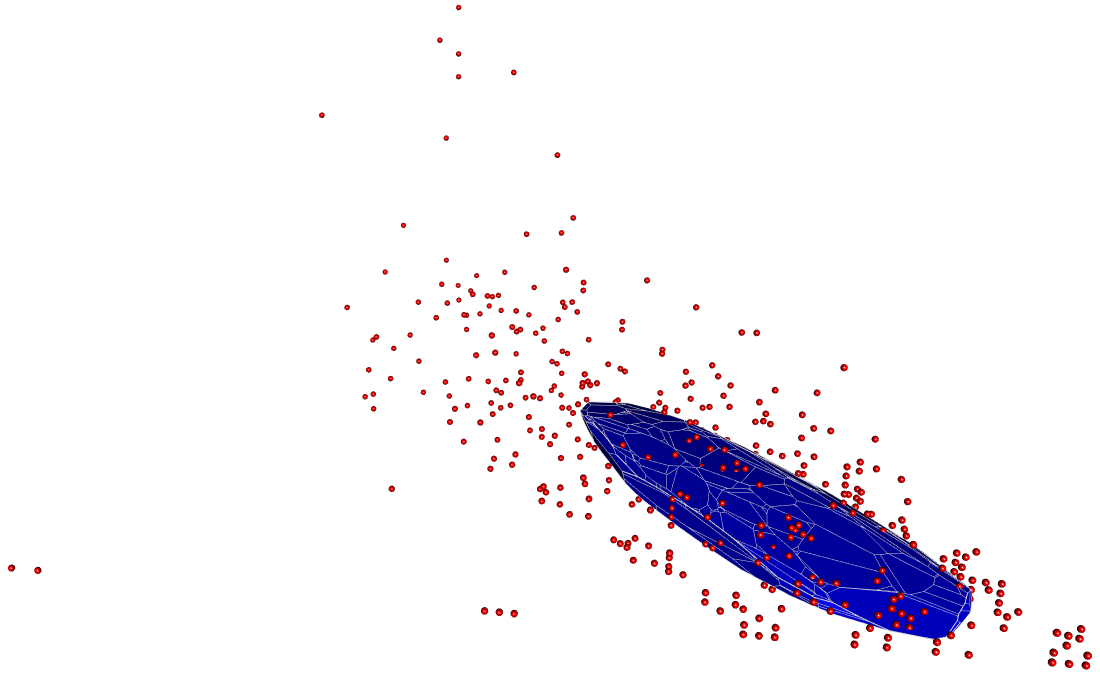}\\
{\small Tukey region for $\kappa = 0.15$ }}}
\fbox{\vbox{\hsize=\www\includegraphics[keepaspectratio=true,width = \ww, trim = 1mm 1mm 1mm 1mm, clip,scale=0.75, page = 1]{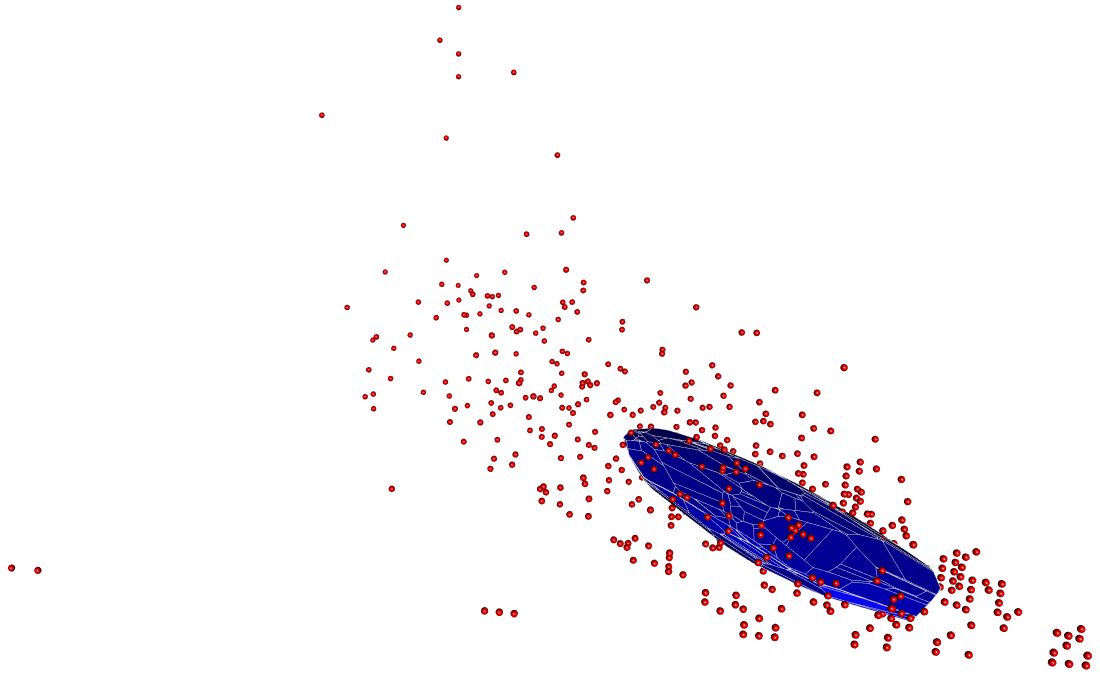}\\
{\small Tukey region for $\kappa = 0.2$ }}}

\fbox{\vbox{\hsize=\www\includegraphics[keepaspectratio=true,width = \ww, trim = 1mm 1mm 1mm 1mm, clip,scale=0.75, page = 1]{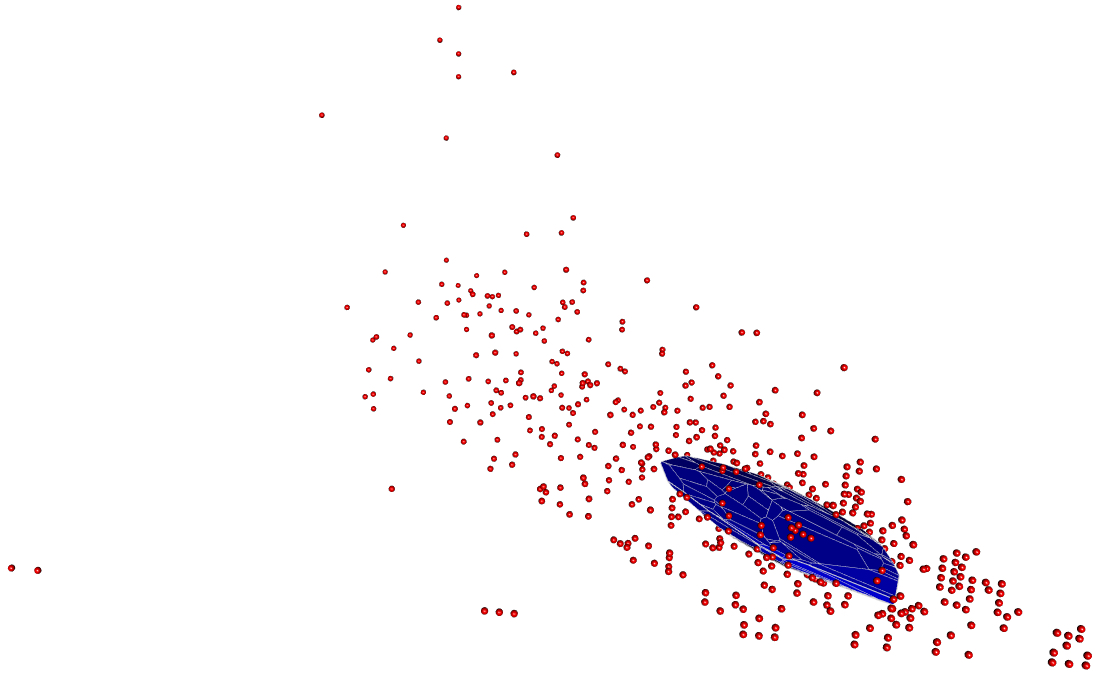}\\
{\small Tukey region for $\kappa = 0.25$ }}}
\fbox{\vbox{\hsize=\www\includegraphics[keepaspectratio=true,width = \ww, trim = 1mm 1mm 1mm 1mm, clip,scale=0.75, page = 1]{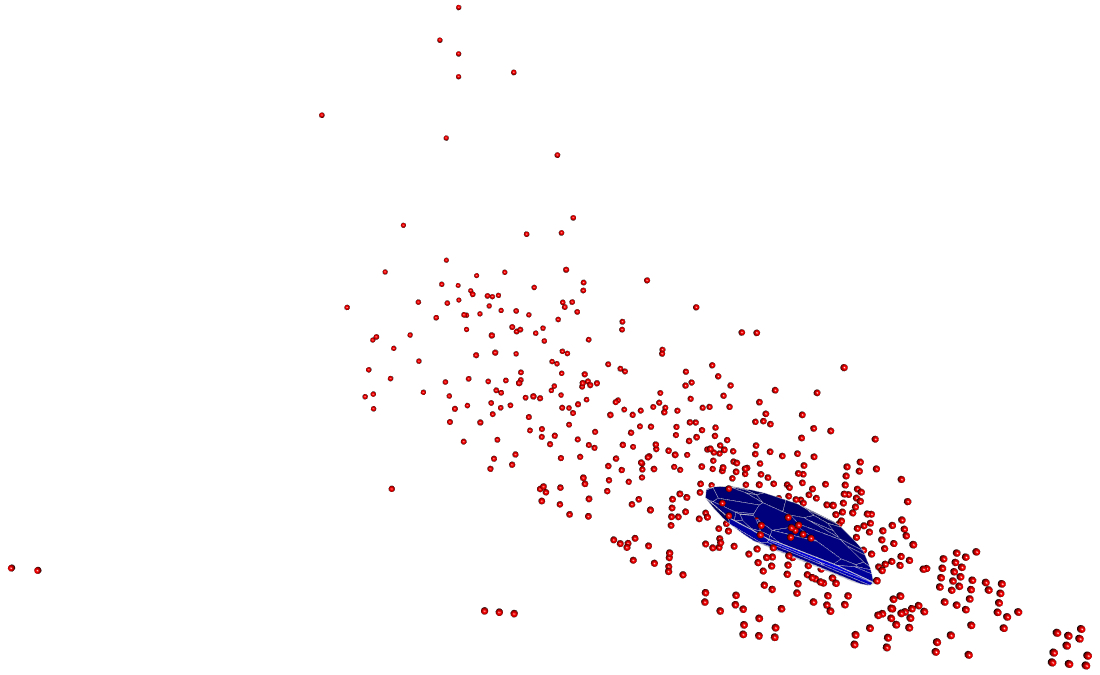}\\
{\small Tukey region for $\kappa = 0.3$ }}}	
		\caption{Tukey $\kappa$-regions for the Blood Transfusion data set.}
		\label{fig:btsample}
	\end{center}
\end{figure}

To illustrate the output of the algorithm, Figure~\ref{fig:btsample} exhibits\footnote{Figures~\ref{fig:btsample}, \ref{fig:ratiosfacets}, \ref{fig:samplemed} and \ref{fig:chemdiab} were generated with the QHULL software written at the Geometry Center, University of Minnesota.}   $\mathcal{D} (\kappa)$ for $\kappa = 0.025,\,0.05,\,0.1,\,0.15,\,0.2,\,0.25,\,0.3$ for the $748$ $3$-dimensional observations of the Blood Transfusion data set, which is downloadable from the UCI Machine Learning Repository~\citep{DuaKT17,YehYT08}. The variables are ``Recency -- months since last donation'', ``Frequency -- total number of donation'', and ``Time -- months since first donation'' \citep[taken from \texttt{R}-package \texttt{ddalpha};][]{ddalphaRpac}. (A fourth variable
``Monetary -- total blood donated in c.c.'' has been removed as it is highly correlated with the third one; see \cite{LiCAL12}.) As we see from Figure~\ref{fig:btsample}, a few contours represent the geometry of the data set \citep[\texttt{R}-package \texttt{rgl} used for visualization;][]{rglRpac}. Further, Table~\ref{tab:cmpBT} indicates that the Algorithm~\ref{alg:1} is faster than Algorithm~\ref{alg:0} due to a smaller number of processed ridges and that most of the found relevant hyperplanes (always coinciding for the two algorithms) are redundant, in particular for deep regions.

\begin{table}[!h]
{\scriptsize
\begin{center}
    \caption{Ratio of computation times and portion of processed ridges by Algorithm~\ref{alg:1} compared to Algorithm~\ref{alg:0}, and ratio ($n^f_\kappa/n_\kappa$) of facets of Tukey regions over the number of relevant hyperplanes  for the Blood Transfusion data set.}
    \label{tab:cmpBT}
    \begin{tabular}{lrrrrrrr} \toprule
    $\kappa$ & 0.025 & 0.05\phantom{0} & 0.1\phantom{000} & 0.15\phantom{00} & 0.2\phantom{000} & 0.25\phantom{00} & 0.3\phantom{0000} \\ 
        \midrule
	Times ratio & 0.034 & 0.1\phantom{00} & 0.27\phantom{00} & 0.45\phantom{00} & 0.61\phantom{00} & 0.76\phantom{00} & 0.87\phantom{000} \\
	Ridges ratio &  0.034 & 0.098 & 0.26\phantom{00} & 0.43\phantom{00} & 0.59\phantom{00} & 0.74\phantom{00} & 0.85\phantom{000} \\
	$n^f_\kappa/n_\kappa$ & 0.097 & 0.029 & 0.0068 & 0.0036 & 0.0022 & 0.0011 & 0.00047 \\
	\bottomrule
	\end{tabular}
\end{center}}
\end{table}

\section{Numerical study}\label{sec:comparisons}

This section presents the results of a simulation study that demonstrates the validity of Algorithm~\ref{alg:1} under diverse data generating distributions and explores its computational performance and algorithmic complexity.
{Results on execution times (Tables~\ref{tab:timealg0}, \ref{tab:timealg1}, \ref{tab:facetsalg1}, \ref{tab:timealg0norm}, \ref{tab:timealg0skewnorm})} are obtained using statistical software \texttt{R} on a Macbook Pro laptop possessing processor Intel(R) Core(TM) i7-4980HQ (2.8 GHz) having 16 GB of physical memory and macOS Sierra (Version 10.12.6) operating system. The validation experiment {as well as the comparison of computation times (Tables~\ref{tab:timealgs} and~\ref{tab:ridges}) have} been conducted on the {\'{E}cole Nationale de la Statistique et de l'Analyse de l'Information}'s computing cluster consisting of the Dell Poweredge M820 servers equipped with 8-kernel processors of type Intel(R) Xeon(R) E5-4627 v2 @ (3.3 GHz) and having 384 GB of physical memory.

\subsection{Validation}\label{ssec:validation}

By construction, Algorithm~\ref{alg:1} yields a region which always includes the true Tukey $\kappa$-region. To explore how close this approximation is, we compare results of Algorithm~\ref{alg:1} with those of the na\"{i}ve Algorithm~\ref{alg:0}, which is known to be correct. We restrict the calculations to those that are feasible regarding Algorithm~\ref{alg:0}, that is, to pairs
  $(n,p)$ for which Algorithm~\ref{alg:0} computes a single Tukey region within less than one hour. As noted above, computation times of Algorithm~\ref{alg:0} are only slightly sensitive to the depth level of the Tukey region. Computation times which are less than one hour, averaged over 10 runs, are given in Table~\ref{tab:timealg0}. The nonempty cells indicate those pairs $(n,p)$  for which the results of the two algorithms regarding relevant hyperplanes will be compared in order to verify the correctness of Algorithm~\ref{alg:1}.

\begin{table}[!ht]
{\scriptsize
\begin{center}
    \caption{Average computation times (in seconds, over 10 runs) of Algorithm~\ref{alg:0}, in case they are less than one hour.}
    \label{tab:timealg0}
    \begin{tabular}{lrrrrrrr}
    \toprule
    $p$\textbackslash & \multicolumn{7}{c}{ $n$}\\
    \cmidrule(r){2-8}
        & \multicolumn{1}{c}{40} & \multicolumn{1}{c}{80} & \multicolumn{1}{c}{160} & \multicolumn{1}{c}{320} & \multicolumn{1}{c}{640} & \multicolumn{1}{c}{1280} & \multicolumn{1}{c}{2560} \\
    \midrule
 		3& 0.003 & 0.03 & 0.2 & 1.4 & 12 & 95 & 792 \\
 		4& 0.04\phantom{0} & 0.6\phantom{0} & 9\phantom{.0} & 152\phantom{.0} & 2\,480 & --- & --- \\
 		5& 0.4\phantom{00} & 12\phantom{.00} & 387\phantom{.0} & ---\phantom{.0} & --- & --- & --- \\
 		6& 3\phantom{.000} & 196\phantom{.00} & ---\phantom{.0} & ---\phantom{.0} & --- & --- & --- \\
 		7& 19\phantom{.000} & 2\,660\phantom{.00} & ---\phantom{.0} & ---\phantom{.0} & --- & --- & --- \\
 		8& 100\phantom{.000} & ---\phantom{.00} & ---\phantom{.0} & ---\phantom{.0} & --- & --- & --- \\
  	    9& 459\phantom{.000} & ---\phantom{.00} & ---\phantom{.0} & ---\phantom{.0} & --- & --- & --- \\
    \bottomrule
    \end{tabular}
\end{center}}
\end{table}

\begin{figure}[!ht]
     \begin{center}
     \footnotesize
\def\ww{.3\textwidth}
\fbox{\vbox{\hsize=.325\textwidth\includegraphics[keepaspectratio=true,width = \ww, trim = 10mm 10mm 10mm 15mm, clip,scale=0.75, page = 1]{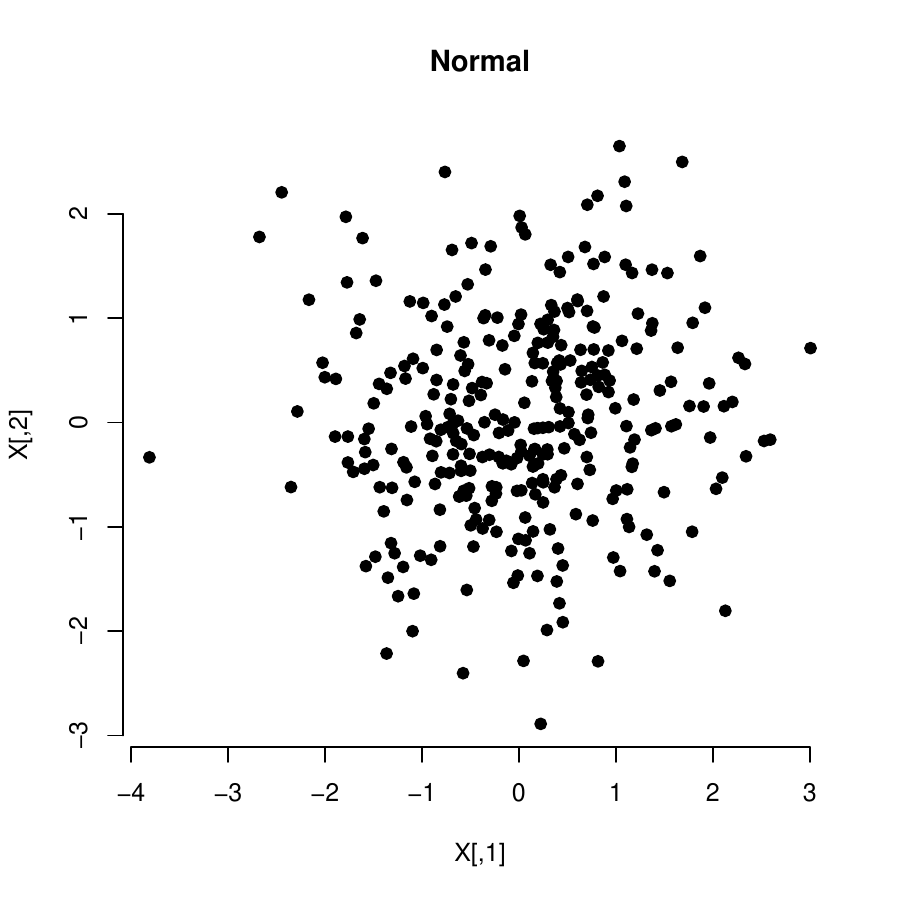}\\
{\small Normal \phantom{j} }}}
\fbox{\vbox{\hsize=.325\textwidth\includegraphics[keepaspectratio=true,width = \ww, trim = 10mm 10mm 10mm 15mm, clip,scale=0.75, page = 2]{pic-samples.pdf}\\
{\small Student t5 \phantom{j} }}}
\fbox{\vbox{\hsize=.325\textwidth\includegraphics[keepaspectratio=true,width = \ww, trim = 10mm 10mm 10mm 15mm, clip,scale=0.75, page = 3]{pic-samples.pdf}\\
{\small Cauchy \phantom{j} }}}
\small
\def\ww{.3\textwidth}
\fbox{\vbox{\hsize=.325\textwidth\includegraphics[keepaspectratio=true,width = \ww, trim = 10mm 10mm 10mm 15mm, clip, scale=0.75, page = 4]{pic-samples.pdf}\\
{\small Uniform in cube \phantom{j} }}}
\fbox{\vbox{\hsize=.325\textwidth\includegraphics[keepaspectratio=true,width = \ww, trim = 10mm 10mm 10mm 15mm, clip, scale=0.75, page = 5]{pic-samples.pdf}\\
{\small Skewed normal \phantom{j} }}}
\fbox{\vbox{\hsize=.325\textwidth\includegraphics[keepaspectratio=true,width = \ww, trim = 10mm 10mm 10mm 15mm, clip, scale=0.75, page = 6]{pic-samples.pdf}\\
{\small Exponential \phantom{j} }}}
	\caption{Data from six distributions; $p=2$,  $n=320$.}
	\label{fig:t1samples}
     \end{center}
\end{figure}

For each of these $21$ pairs $(n,p)$, selected by their computational feasibility, the following experiment is performed. We consider six distributions:
\begin{itemize}
	\item multivariate standard normal distribution;
	\item elliptical Student-$t$ distribution with five degrees of freedom without scaling;
	\item elliptical Cauchy distribution without scaling;
	\item uniform distribution on $[-1,1]^p$;
	\item multivariate skewed-normal distribution with skewness parameter equal to $5$ in the first coordinate according to \cite{AzzaliniC99};
	\item product of independent univariate exponential distributions having parameter $1$.
\end{itemize}
For $p=2$ and $n=320$, a random sample generated from each of them is pictured in Figure~\ref{fig:t1samples}.

\fboxsep0mm

For each of the six distributions we draw $100$ random samples and compute the {relevant} hyperplanes of Tukey regions (stopping after Step~5) with Algorithm~\ref{alg:1} and with Algorithm~\ref{alg:0}. We do this for depth levels selected uniformly from $\{\frac{1}{n},...,\frac{\lfloor 0.35n\rfloor}{n}\}$ (to reduce probability of an empty region). In all $21\times 6 \times 100 = 12600$ cases, the sets of the obtained hyperplanes coincided exactly. (The comparison is facilitated by the fact that each hyperplane is stored as a $p$-tuple of indices of points from the sample and thus can be matched exactly.)

\subsection{Computational performance}\label{ssec:performance}

Further, we measure the time taken for the execution of Algorithm~\ref{alg:1}. Computation times for {several} pairs $(p,n)$ and depth levels are indicated in Table~\ref{tab:timealg1}. As above we restrict the calculations to those of no more than one hour for obtaining relevant {and non-redundant} hyperplanes.
First, the table demonstrates the applicability of the algorithm even for a substantial number of points (up to $5000$)  in dimension three (the still visualizable case) and its capability of computing {relevant} hyperplanes of Tukey regions even in dimension nine as well as calculating the exact shape (facets and vertices) of Tukey regions up to dimension six. Nevertheless, one observes an exponential increase of computation time, which indicates an eventual computational intractability of Algorithm~\ref{alg:1} when the number of observations or the dimension of the data become even larger. Further, one can see that in higher dimensions, while relevant hyperplanes can be found, the QHULL algorithm fails in {determining the non-redundant ones} in a reasonable time. However, in all indicated cases an interior point has been rapidly found in Step~6 of Algorithm~\ref{alg:1}. Note that Algorithm~\ref{alg:1} improves remarkably over Algorithm~\ref{alg:0} for smaller values of $\kappa$. On the other hand when computing deeper regions the ridge-by-ridge strategy can be less efficient and in certain cases even harmful; \textit{cf.} time for $p=6$, $n=40$ and $\kappa=0.3$ in Tables~\ref{tab:timealg0} and~\ref{tab:timealg1}. Such a behavior can be explained by the fact that in this last case a substantial part of $\mathcal{A}$ has to be explored anyway, and the ridge-by-ridge transition takes additional time compared to Algorithm~\ref{alg:0}.

\begin{figure}
\begin{center}
	\includegraphics[width=\textwidth,trim = 0mm 5mm 10mm 20mm,clip]{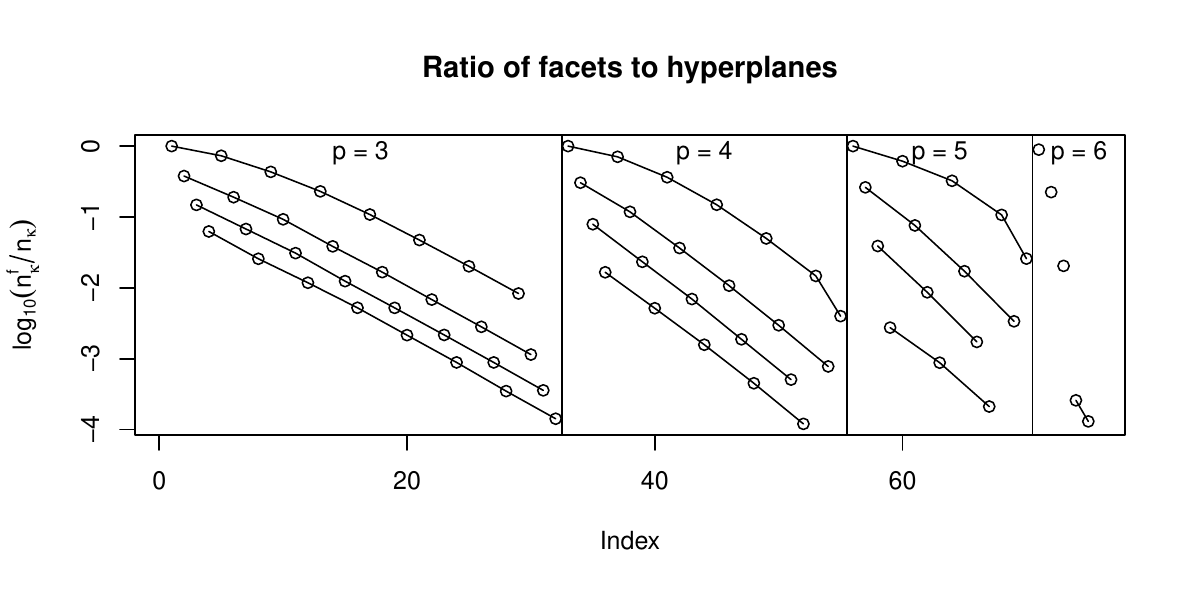}
	\caption{Logarithmized ratio of the number of facets of the Tukey region to the number of relevant hyperplanes, $\log_{10}\frac{n_{\kappa}^f}{n_{\kappa}}$. The index of the points is row-wise($\kappa$ increasing)/column-wise($n$ increasing) for each dimension $p=3,4,5,6$ of Table~\ref{tab:facetsalg1}; \textit{e.g.} sixth point stands for $(n,\kappa)=(80,0.1)$ for $p=3$; lines correspond to rows of Table~\ref{tab:facetsalg1}.}
	\label{fig:ratiosfacets}
\end{center}
\end{figure}

Next, we take a closer look at the numbers of {relevant} hyperplanes and facets of a Tukey region found by Algorithm~\ref{alg:1}. These results are summarized in Table~\ref{tab:facetsalg1}. One observes that in most cases the number of facets is substantially smaller than the number of all relevant hyperplanes (see Figure~\ref{fig:ratiosfacets}); \textit{i.e.} most of the computing time is still utilized unnecessarily. (Note that this refers to the general result by \cite{HallinPS10} and the successive algorithms by \cite{PaindaveineS12a,PaindaveineS12b}). Exceptions are limited to relatively small $n$, and thus do not constitute a computation problem. Further, one can see that the number of facets of the Tukey region is large for middle values of $\kappa$, while it is rather small for both low and high values of the depth. An extreme example of this phenomenon arises with the multivariate Cauchy distribution, where the outer depth regions can be bordered by rather few facets. Also, note that the Tukey median can constitute an arbitrarily tiny polytope having few facets as well.

The results for $\kappa=0.025=1/40$, $n=40$, and $p=6$ attract attention as some of the relevant hyperplanes found -- though all belonging to the convex hull -- are considered redundant. This happens due to precision merging of the QHULL algorithm in higher dimensions. The number of the {relevant} hyperplanes itself, on the other hand, shows exponential growth in $n$ and $p$, which limits the computational feasibility of the Algorithm~\ref{alg:1}.
We further observe that in our experimental settings the upper bound on the facets is never achieved, ranging from close to $0.134$ (in a few rare cases) to $0.0000258$.

\begin{table}
{\scriptsize
\begin{center}
    \caption{Computation times (in seconds) of Algorithm~\ref{alg:1}, taken by finding all relevant hyperplanes (Steps~1 to 5) and by identifying those non-redundant (Steps~6 and 7, in parentheses). For intractable cases, ``t'' indicates reaching the time limit (of one hour) and ``m'' reaching the memory limit (of 16 GB).}
    \label{tab:timealg1}
    \begin{tabular}{lr rrrrrrrr}
    \toprule
    $p$ &$\kappa$ \textbackslash& \multicolumn{8}{c}{Computation times}\\
    \cmidrule(r){3-10}
        & & \multicolumn{1}{c}{40} & \multicolumn{1}{c}{80} & \multicolumn{1}{c}{160} & \multicolumn{1}{c}{320} & \multicolumn{1}{c}{640} & \multicolumn{1}{c}{1280} & \multicolumn{1}{c}{2560} & \multicolumn{1}{c}{5120} \\
    \midrule
3 & 0.025 & 0.0009\phantom{)} & 0.0015\phantom{)} & 0.0069\phantom{)} & 0.042\phantom{0)} & 0.31\phantom{0)}  & 2.2\phantom{)} & 18\phantom{.0)} & 143\phantom{)} \\
 & & (0.0005) & (0.001)\phantom{0} & (0.0022) & (0.0067) & (0.026)  & (0.1)\phantom{} & (0.9) & (7) \\
 & 0.1\phantom{00} & 0.0012\phantom{)} & 0.0071\phantom{)} & 0.044\phantom{0)} & 0.31\phantom{00)} & 2.38\phantom{0)}  & 19\phantom{.0)} & 154\phantom{.0)} & 1\,270\phantom{)} \\
 & & (0.0009) & (0.003)\phantom{0} & (0.0097) & (0.041)\phantom{0} & (0.22)\phantom{0}  & (1.3)\phantom{} & (8)\phantom{.0} & (62) \\
 & 0.2\phantom{00} & 0.0027\phantom{)} & 0.0155\phantom{)} & 0.11\phantom{00)} & 0.75\phantom{00)} & 5.92\phantom{0)}  & 48\phantom{.0)} & 389\phantom{.0)} & 3\,230\phantom{)} \\
 & & (0.0016) & (0.0056) & (0.023)\phantom{0} & (0.1)\phantom{000} & (0.55)\phantom{0}  & (3.4) & (22)\phantom{.0} & (158) \\
 & 0.3\phantom{00} & 0.0039\phantom{)} & 0.023\phantom{0)} & 0.15\phantom{00)} & 1.14\phantom{00)} & 9.15\phantom{0)}  & 75\phantom{.0)} & 602\phantom{.0)} & t\phantom{)} \\
 & & (0.0024) & (0.0074) & (0.032)\phantom{0} & (0.16)\phantom{00} & (0.86)\phantom{0}  & (5.2) & (33)\phantom{.0} & ---\phantom{)} \\[1ex]
4 & 0.025 & 0.001\phantom{0)} & 0.008\phantom{0)} & 0.084\phantom{0)} & 0.9\phantom{000)} & 11\phantom{.000)}  & 162\phantom{.0)} & 2\,440\phantom{.0)} & t\phantom{)} \\
 & & (0.002)\phantom{0} & (0.014)\phantom{0} & (0.078)\phantom{0} & (0.4)\phantom{000} & (3)\phantom{.000}  & (18)\phantom{.0} & (173)\phantom{.0} & ---\phantom{)} \\
 & 0.1\phantom{00} & 0.01\phantom{00)} & 0.094\phantom{0)} & 1.28\phantom{00)} & 17\phantom{.0000)} & 249\phantom{.000)}  & 3\,850\phantom{.0)} & t\phantom{.0)} & ---\phantom{)} \\
 & & (0.017)\phantom{0} & (0.07)\phantom{00} & (0.42)\phantom{00} & (3.6)\phantom{000} & (33)\phantom{.000} & (422)\phantom{.0} & ---\phantom{.0)} & ---\phantom{)} \\
 & 0.2\phantom{00} & 0.027\phantom{0)} & 0.34\phantom{00)} & 4.36\phantom{00)} & 65\phantom{.0000)} & 988\phantom{.000)}  & t\phantom{.0)} & ---\phantom{.0)} & ---\phantom{)} \\
 & & (0.025)\phantom{0} & (0.16)\phantom{00} & (1.29)\phantom{00} & (13)\phantom{.0000} & (132)\phantom{.000}  & ---\phantom{.0)} & ---\phantom{.0)} & ---\phantom{)} \\
 & 0.3\phantom{00} & 0.046\phantom{0)} & 0.62\phantom{00)} & 8.38\phantom{00)} & 121\phantom{.0000)} & 1\,880\phantom{.000)}  & ---\phantom{.0)} & ---\phantom{.0)} & ---\phantom{)} \\
 & & (0.056)\phantom{0} & (0.23)\phantom{00} & (2.28)\phantom{00} & (23)\phantom{.0000} & (294)\phantom{.000}  & ---\phantom{.0)} & ---\phantom{.0)} & ---\phantom{)} \\[1ex]
5 & 0.025 & 0.005\phantom{0)} & 0.05\phantom{00)} & 0.7\phantom{000)} & 14\phantom{.0000)} & 322\phantom{.000)}  & t\phantom{.0)} & ---\phantom{.0)} & ---\phantom{)} \\
 & & (0.02)\phantom{00} & (7.23)\phantom{00} & (28.6)\phantom{000} & (340)\phantom{.0000} & (1\,990)\phantom{.000}  & ---\phantom{.0)} & ---\phantom{.0)} & ---\phantom{)} \\
 & 0.1\phantom{00} & 0.06\phantom{00)} & 1.16\phantom{00)} & 26.2\phantom{000)} & 695\phantom{.0000)} & t\phantom{.000)}  & ---\phantom{.0)} & ---\phantom{.0)} & ---\phantom{)} \\
 & & (1)\phantom{.0000} & (8.25)\phantom{00} & (81.3)\phantom{000} & (590)\phantom{.0000} & ---\phantom{.000)}  & ---\phantom{.0)} & ---\phantom{.0)} & ---\phantom{)} \\
 & 0.2\phantom{00} & 0.24\phantom{00)} & 5.55\phantom{00)} & 149\phantom{.0000)} & t\phantom{.0000)} & ---\phantom{.000)}  & ---\phantom{.0)} & ---\phantom{.0)} & ---\phantom{)} \\
 & & (0.51)\phantom{00} & (5.87)\phantom{00} & (88.5)\phantom{000} & ---\phantom{.0000)} & ---\phantom{.000)}  & ---\phantom{.0)} & ---\phantom{.0)} & ---\phantom{)} \\
 & 0.3\phantom{00} & 0.45\phantom{00)} & 11.9\phantom{000)} & 340\phantom{.0000)} & ---\phantom{.0000)} & ---\phantom{.000)}  & ---\phantom{.0)} & ---\phantom{.0)} & ---\phantom{)} \\
 & & (2.39)\phantom{00} & (6.97)\phantom{00} & (154)\phantom{.0000} & ---\phantom{.0000)} & ---\phantom{.000)}  & ---\phantom{.0)} & ---\phantom{.0)} & ---\phantom{)} \\[1ex]
6 & 0.025 & 0.02\phantom{00)} & 0.3\phantom{000)} & 6\phantom{.0000)} & 201\phantom{.0000)} & m\phantom{.000)}  & ---\phantom{.0)} & ---\phantom{.0)} & ---\phantom{)} \\
 & & (1.8)\phantom{000} & (t)\phantom{.0000} & (t)\phantom{.0000} & (t)\phantom{.0000} & ---\phantom{.000)}  & ---\phantom{.0)} & ---\phantom{.0)} & ---\phantom{)} \\
 & 0.1\phantom{00} & 0.3\phantom{000)} & 12\phantom{.0000)} & 497\phantom{.0000)} & m\phantom{.0000)} & ---\phantom{.000)}  & ---\phantom{.0)} & ---\phantom{.0)} & ---\phantom{)} \\
 & & (1\,630)\phantom{.0000} & (t)\phantom{.0000} & (t)\phantom{.0000} & ---\phantom{.0000)} & ---\phantom{.000)}  & ---\phantom{.0)} & ---\phantom{.0)} & ---\phantom{)} \\
 & 0.2\phantom{00} & 1.9\phantom{000)} & 83\phantom{.0000)} & t\phantom{.0000)} & ---\phantom{.0000)} & ---\phantom{.000)}  & ---\phantom{.0)} & ---\phantom{.0)} & ---\phantom{)} \\
 & & (35)\phantom{.0000} & (t)\phantom{.0000} & ---\phantom{.0000)} & ---\phantom{.0000)} & ---\phantom{.000)}  & ---\phantom{.0)} & ---\phantom{.0)} & ---\phantom{)} \\
 & 0.3\phantom{00} & 4.8\phantom{000)} & 214\phantom{.0000)} & ---\phantom{.0000)} & ---\phantom{.0000)} & ---\phantom{.000)}  & ---\phantom{.0)} & ---\phantom{.0)} & ---\phantom{)} \\
 & & (219)\phantom{.0000} & (1\,480)\phantom{.0000} & ---\phantom{.0000)} & ---\phantom{.0000)} & ---\phantom{.000)}  & ---\phantom{.0)} & ---\phantom{.0)} & ---\phantom{)} \\[1ex]
7 & 0.025 & 0.06\phantom{00)} & 1\phantom{.0000)} & 50\phantom{.0000)} & m\phantom{.0000)} & ---\phantom{.000)}  & ---\phantom{.0)} & ---\phantom{.0)} & ---\phantom{)} \\
 & & (t)\phantom{.0000} & (t)\phantom{.0000} & (t)\phantom{.0000} & ---\phantom{.0000)} & ---\phantom{.000)}  & ---\phantom{.0)} & ---\phantom{.0)} & ---\phantom{)} \\
 & 0.1\phantom{00} & 2\phantom{.0000)} & 101\phantom{.0000)} & m\phantom{.0000)} & ---\phantom{.0000)} & ---\phantom{.000)}  & ---\phantom{.0)} & ---\phantom{.0)} & ---\phantom{)} \\
 & & (t)\phantom{.0000} & (t)\phantom{.0000} & ---\phantom{.0000)} & ---\phantom{.0000)} & ---\phantom{.000)}  & ---\phantom{.0)} & ---\phantom{.0)} & ---\phantom{)} \\
 & 0.2\phantom{00} & 11\phantom{.0000)} & 1\,040\phantom{.0000)} & ---\phantom{.0000)} & ---\phantom{.0000)} & ---\phantom{.000)}  & ---\phantom{.0)} & ---\phantom{.0)} & ---\phantom{)} \\
 & & (t)\phantom{.0000} & (t)\phantom{.0000} & ---\phantom{.0000)} & ---\phantom{.0000)} & ---\phantom{.000)}  & ---\phantom{.0)} & ---\phantom{.0)} & ---\phantom{)} \\[1ex]
8 & 0.025 & 0.2\phantom{000)} & 7\phantom{.0000)} & m\phantom{.0000)} & ---\phantom{.0000)} & ---\phantom{.000)}  & ---\phantom{.0)} & ---\phantom{.0)} & ---\phantom{)} \\
 & & (t)\phantom{.0000} & (t)\phantom{.0000} & ---\phantom{.0000)} & ---\phantom{.0000)} & ---\phantom{.000)}  & ---\phantom{.0)} & ---\phantom{.0)} & ---\phantom{)} \\
 & 0.1\phantom{00} & 6\phantom{.0000)} & 789\phantom{.0000)} & ---\phantom{.0000)} & ---\phantom{.0000)} & ---\phantom{.000)}  & ---\phantom{.0)} & ---\phantom{.0)} & ---\phantom{)} \\
 & & (t)\phantom{.0000} & (t)\phantom{.0000} & ---\phantom{.0000)} & ---\phantom{.0000)} & ---\phantom{.000)}  & ---\phantom{.0)} & ---\phantom{.0)} & ---\phantom{)} \\
 & 0.2\phantom{00} & 55\phantom{.0000)} & m\phantom{.0000)} & ---\phantom{.0000)} & ---\phantom{.0000)} & ---\phantom{.000)}  & ---\phantom{.0)} & ---\phantom{.0)} & ---\phantom{)} \\
 & & (t)\phantom{.0000} & ---\phantom{.0000)} & ---\phantom{.0000)} & ---\phantom{.0000)} & ---\phantom{.000)}  & ---\phantom{.0)} & ---\phantom{.0)} & ---\phantom{)} \\[1ex]
9 & 0.025 & 0.6\phantom{000)} & 30\phantom{.0000)} & m\phantom{.0000)} & ---\phantom{.0000)} & ---\phantom{.000)}  & ---\phantom{.0)} & ---\phantom{.0)} & ---\phantom{)} \\
 & & (t)\phantom{.0000} & (t)\phantom{.0000} & ---\phantom{.0000)} & ---\phantom{.0000)} & ---\phantom{.000)}  & ---\phantom{.0)} & ---\phantom{.0)} & ---\phantom{)} \\
 & 0.1\phantom{00} & 25\phantom{.0000)} & m\phantom{.0000)} & ---\phantom{.0000)} & ---\phantom{.0000)} & ---\phantom{.000)}  & ---\phantom{.0)} & ---\phantom{.0)} & ---\phantom{)} \\
 & & (t)\phantom{.0000} & ---\phantom{.0000)} & ---\phantom{.0000)} & ---\phantom{.0000)} & ---\phantom{.000)}  & ---\phantom{.0)} & ---\phantom{.0)} & ---\phantom{)} \\
 & 0.2\phantom{00}  & 259\phantom{.0000)} & ---\phantom{.0000)} & ---\phantom{.0000)} & ---\phantom{.0000)} & ---\phantom{.000)}  & ---\phantom{.0)} & ---\phantom{.0)} & ---\phantom{)} \\
 & & (t)\phantom{.0000} & ---\phantom{.0000)} & ---\phantom{.0000)} & ---\phantom{.0000)} & ---\phantom{.000)} & ---\phantom{.0)} & ---\phantom{.0)} & ---\phantom{)} \\[1ex]
    \bottomrule
    \end{tabular}
\end{center}}
\end{table}

\begin{table}
{\scriptsize
\begin{center}
    \caption{Number of relevant hyperplanes and those non-redundant (in parentheses) delivered by Algorithm~\ref{alg:1}. For intractable cases, ``t'' indicates reaching the time limit (of one hour) and ``m'' reaching the memory limit (of 16 GB).}
    \label{tab:facetsalg1}
    \begin{tabular}{lr rrrrrrrr}
    \toprule
    $p$ &$\kappa$ \textbackslash& \multicolumn{8}{c}{Number of hyperplanes}\\
    \cmidrule(r){3-10}
        &      & \multicolumn{1}{c}{40} & \multicolumn{1}{c}{80} & \multicolumn{1}{c}{160} & \multicolumn{1}{c}{320} & \multicolumn{1}{c}{640} & \multicolumn{1}{c}{1280} & \multicolumn{1}{c}{2560} & \multicolumn{1}{c}{5120} \\
    \midrule
3 & 0.025 & 29\phantom{)} & 95\phantom{)} & 369\phantom{)} & 1\,380\phantom{)} & 5\,280\phantom{)} & 20\,400\phantom{)} & 82\,000\phantom{)} & 326\,000\phantom{)} \\
 & & (29) & (69) & (160) & (318) & (571) & (965) & (1\,650) & (2\,710) \\
 & 0.1\phantom{00} & 185\phantom{)} & 753\phantom{)} & 2\,890\phantom{)} & 11\,500\phantom{)} & 46\,000\phantom{)} & 184\,000\phantom{)} & 732\,000\phantom{)} & 2\,930\,000\phantom{)} \\
 & & (70) & (143) & (268) & (441) & (765) & (1\,250) & (2\,060) & (3\,350) \\
 & 0.2\phantom{00} & 449\phantom{)} & 1\,830\phantom{)} & 7\,340\phantom{)} & 29\,100\phantom{)} & 116\,000\phantom{)} & 467\,000\phantom{)} & 1\,860\,000\phantom{)} & 7\,460\,000\phantom{)} \\
 & & (66) & (124) & (227) & (362) & (606) & (1\,010) & (1\,650) & (2\,670) \\
 & 0.3\phantom{00} & 677\phantom{)} & 2\,720\phantom{)} & 11\,200\phantom{)} & 44\,700\phantom{)} & 181\,000\phantom{)} & 719\,000\phantom{)} & 2\,880\,000\phantom{)} & t\phantom{)} \\
 & & (42) & (70) & (132) & (234) & (388) & (637) & (1\,000) & ---\phantom{)} \\[1ex]
4 & 0.025 & 100\phantom{)} & 510\phantom{)} & 3\,250\phantom{)} & 21\,900\phantom{)} & 153\,000\phantom{)} & 1\,160\,000\phantom{)} & 9\,020\,000\phantom{)} & t\phantom{)} \\
 & & (100) & (360) & (1\,180) & (3\,250) & (7\,620) & (17\,100) & (36\,000) & ---\phantom{)} \\
 & 0.1\phantom{00} & 1\,130\phantom{)} & 8\,180\phantom{)} & 61\,600\phantom{)} & 482\,000\phantom{)} & 3\,780\,000\phantom{)} & 30\,100\,000\phantom{)} & t\phantom{)} & ---\phantom{)} \\
 & & (343) & (966) & (2\,250) & (5\,170) & (11\,200) & (23\,400) & ---\phantom{)} & ---\phantom{)} \\
 & 0.2\phantom{00} & 3\,810\phantom{)} & 30\,000\phantom{)} & 236\,000\phantom{)} & 1\,890\,000\phantom{)} & 15\,100\,000\phantom{)} & t\phantom{)} & ---\phantom{)} & ---\phantom{)} \\
 & & (301) & (700) & (1\,640) & (3\,550) & (7\,680) & ---\phantom{)} & ---\phantom{)} & ---\phantom{)} \\
 & 0.3\phantom{00} & 6\,750\phantom{)} & 55\,500\phantom{)} & 448\,000\phantom{)} & 3\,580\,000\phantom{)} & 28\,800\,000\phantom{)} & ---\phantom{)} & ---\phantom{)} & ---\phantom{)} \\
 & & (112) & (286) & (705) & (1\,610) & (3\,470) & ---\phantom{)} & ---\phantom{)} & ---\phantom{)} \\[1ex]
5 & 0.025 & 328\phantom{)} & 2\,550\phantom{)} & 24\,400\phantom{)} & 272\,000\phantom{)} & 3\,550\,000\phantom{)} & t\phantom{)} & ---\phantom{)} & ---\phantom{)} \\
 & & (328) & (1\,570) & (7\,920) & (29\,100) & (91\,700) & ---\phantom{)} & ---\phantom{)} & ---\phantom{)} \\
 & 0.1\phantom{00} & 5\,850\phantom{)} & 76\,100\phantom{)} & 1\,060\,000\phantom{)} & 15\,700\,000\phantom{)} & t\phantom{)} & ---\phantom{)} & ---\phantom{)} & ---\phantom{)} \\
 & & (1\,530) & (5\,780) & (18\,200) & (52\,800) & ---\phantom{)} & ---\phantom{)} & ---\phantom{)} & ---\phantom{)} \\
 & 0.2\phantom{00} & 26\,100\phantom{)} & 396\,000\phantom{)} & 6\,140\,000\phantom{)} & t\phantom{)} & ---\phantom{)} & ---\phantom{)} & ---\phantom{)} & ---\phantom{)} \\
 & & (1\,020) & (3\,430) & (10\,600) & ---\phantom{)} & ---\phantom{)} & ---\phantom{)} & ---\phantom{)} & ---\phantom{)} \\
 & 0.3\phantom{00} & 52\,200\phantom{)} & 863\,000\phantom{)} & 13\,900\,000\phantom{)} & ---\phantom{)} & ---\phantom{)} & ---\phantom{)} & ---\phantom{)} & ---\phantom{)} \\
 & & (143) & (758) & (2\,950) & ---\phantom{)} & ---\phantom{)} & ---\phantom{)} & ---\phantom{)} & ---\phantom{)} \\[1ex]
6 & 0.025 & 968\phantom{)} & 11\,900\phantom{)} & 169\,000\phantom{)} & 3\,110\,000\phantom{)} & m\phantom{)} & ---\phantom{)} & ---\phantom{)} & ---\phantom{)} \\
 & & (864) & (t) & (t) & (t) & ---\phantom{)} & ---\phantom{)} & ---\phantom{)} & ---\phantom{)} \\
 & 0.1\phantom{00} & 25\,200\phantom{)} & 603\,000\phantom{)} & 15\,200\,000\phantom{)} & m\phantom{)} & ---\phantom{)} & ---\phantom{)} & ---\phantom{)} & ---\phantom{)} \\
 & & (5\,650) & (t) & (t) & ---\phantom{)} & ---\phantom{)} & ---\phantom{)} & ---\phantom{)} & ---\phantom{)} \\
 & 0.2\phantom{00} & 143\,000\phantom{)} & 4\,320\,000\phantom{)} & t\phantom{)} & ---\phantom{)} & ---\phantom{)} & ---\phantom{)} & ---\phantom{)} & ---\phantom{)} \\
 & & (2\,930) & (t) & ---\phantom{)} & ---\phantom{)} & ---\phantom{)} & ---\phantom{)} & ---\phantom{)} & ---\phantom{)} \\
 & 0.3\phantom{00} & 327\,000\phantom{)} & 11\,000\,000\phantom{)} & ---\phantom{)} & ---\phantom{)} & ---\phantom{)} & ---\phantom{)} & ---\phantom{)} & ---\phantom{)} \\
 & & (85) & (1\,430) & ---\phantom{)} & ---\phantom{)} & ---\phantom{)} & ---\phantom{)} & ---\phantom{)} & ---\phantom{)} \\[1ex]
7 & 0.025 & 2\,630\phantom{)} & 48\,400\phantom{)} & 1\,070\,000\phantom{)} & m\phantom{)} & ---\phantom{)} & ---\phantom{)} & ---\phantom{)} & ---\phantom{)} \\
 & & (t) & (t) & (t) & ---\phantom{)} & ---\phantom{)} & ---\phantom{)} & ---\phantom{)} & ---\phantom{)} \\
 & 0.1\phantom{00} & 95\,300\phantom{)} & 3\,980\,000\phantom{)} & m\phantom{)} & ---\phantom{)} & ---\phantom{)} & ---\phantom{)} & ---\phantom{)} & ---\phantom{)} \\
 & & (t) & (t) & ---\phantom{)} & ---\phantom{)} & ---\phantom{)} & ---\phantom{)} & ---\phantom{)} & ---\phantom{)} \\
 & 0.2\phantom{00} & 678\,000\phantom{)} & 39\,200\,000\phantom{)} & ---\phantom{)} & ---\phantom{)} & ---\phantom{)} & ---\phantom{)} & ---\phantom{)} & ---\phantom{)} \\
 & & (t) & (t) & ---\phantom{)} & ---\phantom{)} & ---\phantom{)} & ---\phantom{)} & ---\phantom{)} & ---\phantom{)} \\[1ex]
8 & 0.025 & 6\,930\phantom{)} & 189\,000\phantom{)} & m\phantom{)} & ---\phantom{)} & ---\phantom{)} & ---\phantom{)} & ---\phantom{)} & ---\phantom{)} \\
 & & (t) & (t) & ---\phantom{)} & ---\phantom{)} & ---\phantom{)} & ---\phantom{)} & ---\phantom{)} & ---\phantom{)} \\
 & 0.1\phantom{00} & 320\,000\phantom{)} & 24\,400\,000\phantom{)} & ---\phantom{)} & ---\phantom{)} & ---\phantom{)} & ---\phantom{)} & ---\phantom{)} & ---\phantom{)} \\
 & & (t) & (t) & ---\phantom{)} & ---\phantom{)} & ---\phantom{)} & ---\phantom{)} & ---\phantom{)} & ---\phantom{)} \\
 & 0.2\phantom{00} & 2\,790\,000\phantom{)} & m\phantom{)} & ---\phantom{)} & ---\phantom{)} & ---\phantom{)} & ---\phantom{)} & ---\phantom{)} & ---\phantom{)} \\
 & & (t) & ---\phantom{)} & ---\phantom{)} & ---\phantom{)} & ---\phantom{)} & ---\phantom{)} & ---\phantom{)} & ---\phantom{)} \\[1ex]
9 & 0.025 & 18\,800\phantom{)} & 670\,000\phantom{)} & m\phantom{)} & ---\phantom{)} & ---\phantom{)} & ---\phantom{)} & ---\phantom{)} & ---\phantom{)} \\
 & & (t) & (t) & ---\phantom{)} & ---\phantom{)} & ---\phantom{)} & ---\phantom{)} & ---\phantom{)} & ---\phantom{)} \\
 & 0.1\phantom{00} & 1\,040\,000\phantom{)} & m\phantom{)} & ---\phantom{)} & ---\phantom{)} & ---\phantom{)} & ---\phantom{)} & ---\phantom{)} & ---\phantom{)} \\
 & & (t) & ---\phantom{)} & ---\phantom{)} & ---\phantom{)} & ---\phantom{)} & ---\phantom{)} & ---\phantom{)} & ---\phantom{)} \\
 & 0.2\phantom{00} & 10\,300\,000\phantom{)} & ---\phantom{)} & ---\phantom{)} & ---\phantom{)} & ---\phantom{)} & ---\phantom{)} & ---\phantom{)} & ---\phantom{)} \\
 & & (t) & ---\phantom{)} & ---\phantom{)} & ---\phantom{)} & ---\phantom{)} & ---\phantom{)} & ---\phantom{)} & ---\phantom{)} \\[1ex]
    \bottomrule
    \end{tabular}
\end{center}}
\end{table}

\subsection{Time and complexity comparison}\label{ssec:time}

Algorithm~\ref{alg:1} has time complexity $O\bigl(n^p\log(n)\bigr)$. For the algorithm developed by \cite{PaindaveineS12a}, henceforth \emph{Algorithm~PSa}, time complexity amounts to $O(n^{p+1})$. The authors conjecture that for a random $\kappa\in(0,0.5)$ the average time complexity is $O(n^p)$. The simulation study of the procedure given in \cite{PaindaveineS12b}, henceforth \emph{Algorithm~PSb}, points to a similar complexity. As Algorithm~\ref{alg:1} is of a different nature than PSa and PSb, and the complexities disregard $\kappa$, they cannot be directly compared.

To gain further insights, we compare (average) computation times of the three algorithms. Table~\ref{tab:timealgs} indicates how much faster Algorithm~\ref{alg:1} is than the two competitors (while always yielding the same relevant hyperplanes). At first sight, Algorithm~\ref{alg:1} is faster by several orders. However, the speed gain decreases with $n$, suggesting that Algorithm~\ref{alg:1} has higher time complexity. To explain this last observation, see Table~\ref{tab:ridges}. First, we indicate the (average) portion of ridges processed by Algorithm~\ref{alg:1} compared to Algorithm~\ref{alg:0}, where it is constant and equal to ${n \choose p - 1}$. These numbers suggest that Algorithm~\ref{alg:1} can be much more efficient than Algorithm~\ref{alg:0}, as it normally processes fewer ridges and this difference increases with decreasing $\kappa$. Second, we report the ratio of the number of the relevant hyperplanes found by Algorithm~PSa to the number of ridges processed by Algorithm~\ref{alg:1}. As the number of relevant hyperplanes corresponds to the number of cones processed by Algorithm~PSa, this ratio indicates that the number of the elements (ridges for Algorithm~\ref{alg:1}, cones for Algorithm~PSa) processed by both Algorithm~\ref{alg:1} and Algorithm~PSa  is of the same order. Nevertheless, processing one element has time complexity $O\bigl(n\log(n)\bigr)$ in Algorithm~\ref{alg:1} and $O(n)$ in Algorithm~PSa. For Algorithm~PSb a similar logic applies. {On the other hand}, Table~\ref{tab:timealgs} shows that Algorithm~\ref{alg:1} is faster even for large $n$. This is due to the simplicity of its operations and the lightness of the involved structures. In particular, neither post-optimization nor the QHULL algorithm are used {when searching for relevant hyperplanes}, and a ridge is saved as a simple combination of data points.

\begin{table}[!t]
{\scriptsize
\begin{center}
    \caption{Ratios of average computation times of Algorithms~PSa and~PSb over  that of Algorithm~\ref{alg:1}. ($10$ repetitions, standard multivariate normal distribution, implementation in \texttt{R}-package \texttt{modQR} \citep{modQRRpac}.)}
    \label{tab:timealgs}
    \begin{tabular}{lrc rrrrrrrr}
    \toprule
    $p$ & $\kappa$ & Algorithm \textbackslash& \multicolumn{8}{c}{Computation time ratios}\\
    \cmidrule(r){4-11}
        & & & \multicolumn{1}{c}{40} & \multicolumn{1}{c}{80} & \multicolumn{1}{c}{160} & \multicolumn{1}{c}{320} & \multicolumn{1}{c}{640} & \multicolumn{1}{c}{1280} & \multicolumn{1}{c}{2560} & \multicolumn{1}{c}{5120} \\
    \midrule
3 & 0.025 & PSa & 4\,920 & 8\,340 & 6\,490 & 2\,980 & 2\,060 & 676 & 268 & 149 \\
 & & PSb & 3\,760 & 7\,240 & 7\,230 & 4\,100 & 2\,370 & 600 & 269 & 157 \\
 & 0.1\phantom{00} & PSa & 23\,500 & 20\,100 & 9\,670 & 5\,090 & 2\,190 & 1\,000 & 420 & --- \\
 & & PSb & 22\,900 & 22\,000 & 17\,300 & 6\,300 & 2\,540 & 982 & 361 & --- \\
 & 0.2\phantom{00} & PSa & 25\,300 & 25\,400 & 11\,100 & 4\,410 & 2\,450 & 924 & --- & --- \\
 & & PSb & 26\,100 & 29\,200 & 17\,800 & 5\,950 & 2\,080 & 784 & --- & --- \\
 & 0.3\phantom{00} & PSa & 21\,200 & 27\,500 & 11\,400 & 4\,910 & 2\,400 & 1\,030 & --- & --- \\
 & & PSb & 23\,400 & 35\,500 & 11\,900 & 5\,050 & 2\,240 & 883 & --- & --- \\[1ex]
4 & 0.025 & PSa & 16\,500 & 13\,700 & 6\,320 & 3\,510 & 1\,830 & 1\,620 & --- & --- \\
 & & PSb & 7\,580 & 11\,700 & 9\,860 & 2\,860 & 1\,760 & 843 & --- & --- \\
 & 0.1\phantom{00} & PSa & 66\,700 & 22\,300 & 16\,200 & 6\,960 & --- & --- & --- & --- \\
 & & PSb & 57\,500 & 28\,300 & 19\,600 & 4\,850 & --- & --- & --- & --- \\
 & 0.2\phantom{00} & PSa & 57\,300 & 36\,000 & 19\,000 & --- & --- & --- & --- & --- \\
 & & PSb & 57\,500 & 35\,400 & 10\,100 & --- & --- & --- & --- & --- \\
 & 0.3\phantom{00} & PSa & 55\,700 & 40\,600 & 17\,400 & --- & --- & --- & --- & --- \\
 & & PSb & 62\,100 & 40\,800 & 12\,600 & --- & --- & --- & --- & --- \\[1ex]
    \bottomrule
    \end{tabular}
\end{center}}
\end{table}

\begin{table}[!h]
{\scriptsize
\begin{center}
    \caption{Portion of ridges processed by Algorithm~\ref{alg:1} compared to Algorithm~\ref{alg:0}, and (in parentheses) the ratio of the number of relevant hyperplanes found by Algorithm~PSa to the number of ridges processed by Algorithm~\ref{alg:1}.}
    \label{tab:ridges}
    \begin{tabular}{lc rrrrrrrr}
    \toprule
    $p$ & $\kappa$ & \multicolumn{8}{c}{Portion of processed ridges (non-redundant hyperplanes)}\\
    \cmidrule(r){3-10}
        & & \multicolumn{1}{c}{40} & \multicolumn{1}{c}{80} & \multicolumn{1}{c}{160} & \multicolumn{1}{c}{320} & \multicolumn{1}{c}{640} & \multicolumn{1}{c}{1280} & \multicolumn{1}{c}{2560} & \multicolumn{1}{c}{5120}  \\
    \midrule
3 & 0.025 & 0.059\phantom{)} & 0.044\phantom{)} & 0.035\phantom{0)} & 0.029\phantom{0)} & 0.026\phantom{0)} & 0.025\phantom{)} & 0.023\phantom{)} & 0.023\phantom{)} \\
 & & (0.68)\phantom{0} & (0.92)\phantom{0} & (1.64)\phantom{00} & (2.18)\phantom{00} & (2.64)\phantom{00} & (2.9)\phantom{00} & (3.08)\phantom{0} & (3.26)\phantom{0} \\
 & 0.1\phantom{00} & 0.29\phantom{0)} & 0.26\phantom{0)} & 0.23\phantom{00)} & 0.22\phantom{00)} & 0.21\phantom{00)} & 0.2\phantom{00)} & 0.2\phantom{00)} & ---\phantom{.000)} \\
 & & (1.69)\phantom{0} & (2.28)\phantom{0} & (2.68)\phantom{00} & (2.98)\phantom{00} & (3.18)\phantom{00} & (3.29)\phantom{0} & (3.38)\phantom{0} & ---\phantom{.000)} \\
 & 0.2\phantom{00} & 0.62\phantom{0)} & 0.58\phantom{0)} & 0.55\phantom{00)} & 0.53\phantom{00)} & 0.52\phantom{00)} & 0.51\phantom{0)} & ---\phantom{.000)} & ---\phantom{.000)} \\
 & & (2.41)\phantom{0} & (2.76)\phantom{0} & (3)\phantom{.0000} & (3.23)\phantom{00} & (3.3)\phantom{000} & (3.35)\phantom{0} & ---\phantom{.000)} & ---\phantom{.000)} \\
 & 0.3\phantom{00} & 0.87\phantom{0)} & 0.84\phantom{0)} & 0.81\phantom{00)} & 0.79\phantom{00)} & 0.78\phantom{00)} & 0.77\phantom{0)} & ---\phantom{.000)} & ---\phantom{.000)} \\
 & & (2.77)\phantom{0} & (2.99)\phantom{0} & (3.13)\phantom{00} & (3.27)\phantom{00} & (3.35)\phantom{00} & (3.41)\phantom{0} & ---\phantom{.000)} & ---\phantom{.000)} \\[1ex]
4 & 0.025 & 0.02\phantom{0)} & 0.012\phantom{)} & 0.0075\phantom{)} & 0.0055\phantom{)} & 0.0045\phantom{)} & 0.004\phantom{)} & ---\phantom{.000)} & ---\phantom{.000)} \\
 & & (0.47)\phantom{0} & (0.72)\phantom{0} & (1.17)\phantom{00} & (1.89)\phantom{00} & (2.51)\phantom{00} & (2.98)\phantom{0} & ---\phantom{.000)} & ---\phantom{.000)} \\
 & 0.1\phantom{00} & 0.18\phantom{0)} & 0.14\phantom{0)} & 0.12\phantom{00)} & 0.1\phantom{000)} & ---\phantom{.0000)} & ---\phantom{.000)} & ---\phantom{.000)} & ---\phantom{.000)} \\
 & & (1.25)\phantom{0} & (2.01)\phantom{0} & (2.63)\phantom{00} & (3.09)\phantom{00} & ---\phantom{.0000)} & ---\phantom{.000)} & ---\phantom{.000)} & ---\phantom{.000)} \\
 & 0.2\phantom{00} & 0.52\phantom{0)} & 0.45\phantom{0)} & 0.41\phantom{00)} & ---\phantom{.0000)} & ---\phantom{.0000)} & ---\phantom{.000)} & ---\phantom{.000)} & ---\phantom{.000)} \\
 & & (2.09)\phantom{0} & (2.7)\phantom{00} & (3.2)\phantom{000} & ---\phantom{.0000)} & ---\phantom{.0000)} & ---\phantom{.000)} & ---\phantom{.000)} & ---\phantom{.000)} \\
 & 0.3\phantom{00} & 0.84\phantom{0)} & 0.78\phantom{0)} & 0.74\phantom{00)} & ---\phantom{.0000)} & ---\phantom{.0000)} & ---\phantom{.000)} & ---\phantom{.000)} & ---\phantom{.000)} \\
 & & (2.75)\phantom{0} & (3.18)\phantom{0} & (3.46)\phantom{00} & ---\phantom{.0000)} & ---\phantom{.0000)} & ---\phantom{.000)} & ---\phantom{.000)} & ---\phantom{.000)} \\[1ex]
    \bottomrule
    \end{tabular}
\end{center}}
\end{table}

\section{Computing the Tukey median}\label{sec:median}

The \emph{Tukey median} is the gravity center of the Tukey median set, that is, the Tukey region having maximum depth.
With the above algorithms we are able to compute Tukey regions at any given depth level.
However the maximum Tukey depth, $\kappa^*$, depends on the data and has to be determined from them.
For data in general position it is known \citep{LiuLZ16} that
\begin{equation}\label{eq_maxdepth}
 \frac 1n \left\lceil\frac{n}{p+1}\right\rceil
 \le \kappa^* \le
 \frac 1n \left\lfloor \frac{n-p+2}{2}\right\rfloor\,.
\end{equation}
In two-dimensional space, \cite{RousseeuwR98}, hereafter RR, have developed an algorithm, which has time complexity $O\bigl(n^2 \log^2(n)\bigr)$. Their algorithm employs a bisection strategy that starts with a lower and an upper bound on $\kappa^*$ and updates these bounds until they coincide.
In updating, RR calculate the mean $\overline \kappa$ of the two active bounds and check whether $\mathcal{D}(\overline\kappa)$ exists. If yes, the lower bound is changed to $\overline\kappa$; if no, the upper bound becomes $\overline\kappa$.
The RR bisection approach extends easily to data of dimension greater than $2$.
With our new Algorithm~\ref{alg:1} at hand, it is natural to investigate its possible use and accuracy in computing higher-dimensional Tukey medians.

In searching for the maximum level of Tukey depth, we  employ a {modified} bisection strategy. Given a Tukey region $\mathcal {D}(\kappa)$ at some level $\kappa$, we consider its barycenter.
If the region is nonempty, its barycenter is more central than its boundary points, and thus may have a larger depth value.
Observe that computing the depth of a single point is computationally cheaper (\textit{i.e.} has  {lower} time complexity) than computing a trimmed region. Hence, after having checked  that at a given depth level $\kappa$ the region $\mathcal {D}(\kappa)$  is nonempty, we may further compute the barycenter of $\mathcal {D}(\kappa)$ to possibly come closer to the depth maximum. This motivates us to construct the following algorithm. In the algorithm, ${\it\texttt{Tukey\_depth}}(\bmx\,;\,\x_1, \cdots ,\x_n)$ stands for any procedure that computes the Tukey depth of a point $\bmx\in\mathbb{R}^p$ w.r.t.  data $\x_1, \cdots ,\x_n\subset\mathbb{R}^p$, like those in \cite{DyckerhoffM16} or \cite{Liu17}. Further, ${\it\texttt{Alg1}}(\bmx\,;\,\x_1, \cdots ,\x_n; \kappa; \epsilon)$ signifies a Tukey region resulting from Algorithm~\ref{alg:1}.

\begin{algorithm}[Algorithm for computing the Tukey median] \label{alg:median} \quad
\begin{enumerate}
	\item[] {\bf Input:} $\x_1, \cdots ,\x_n \subset \mathbb{R}^p$, $\epsilon$.
	\item[] {\bf Step 1.} {\bf Initialize bounds on $\kappa^*$:}
	\begin{enumerate}
    		\item[(a)] Compute $\x_0 = \bigl(\text{med}(\x_{11}, \cdots ,\x_{n1}), \text{med}(\x_{12}, \cdots ,\x_{n2}), \cdots ,\text{med}(\x_{1d}, \cdots ,\x_{nd})\bigl)^{\top}$.
    		\item[(b)] Compute $d_0=\texttt{Tukey\_depth}(\x_0\,;\, \x_1, \cdots ,\x_n)$.
    		\item[(c)] Set $\kappa_{low} = \max\left\{\frac 1n \lceil \frac{n}{p + 1}\rceil, d_0\right\}$, $\kappa_{up} = \frac 1n \lfloor \frac{n - p + 2}{2} \rfloor + \frac 1n$.
    	\end{enumerate}
    	\item[] {\bf Step 2.} {\bf Update bounds:}
    	\begin{enumerate}
    		\item[] Let $\overline\kappa= \frac 1n \lfloor\frac{n(\kappa_{low} + \kappa_{up})}{2}\rfloor$, and compute the region $\mathcal{D}(\overline\kappa)=\texttt{Alg1}(\x_1, \cdots ,\x_n\,;\,\overline\kappa\,;\,\epsilon)$.
\item    If $\mathcal{D}(\overline\kappa)$ does not exist (that is, \emph{Algorithm~\ref{alg:1}} stops at its \emph{Step~6}), then set\\
     $\kappa_{up}= \overline\kappa$,
\item   If $\mathcal{D}(\overline\kappa)$  exists,\\
 then calculate the barycenter $\mathbf{c}$ of $\mathcal{D}(\overline\kappa)$ and set \\
 $\kappa_{low} = \texttt{Tukey\_depth}(\boldsymbol{c}\,;\,\x_1, \cdots ,\x_n)$.
\item If $\kappa_{low} < \kappa_{up} - \frac{1}{n}$, then repeat {\bf Step~2}, else stop.
    	\end{enumerate}
    	\item[] {\bf Output:} $\texttt{Alg1}(\x_1, \cdots ,\x_n\,;\,\kappa_{low}\,;\,\epsilon)$.
\end{enumerate}
\end{algorithm}

{Step~1c initializes the lower bound either according to (\ref{eq_maxdepth}) or with}
 the depth of the coordinate-wise median, which is cheaply computed and often has a rather high depth value.
In Step~2b, the depth of the barycenter of the Tukey region is computed (in case it exists) to obtain a higher value for the lower bound, which substantially contributes to the speed of Algorithm~\ref{alg:median}.

We compare our Algorithm~\ref{alg:median} with the  bisection approach of RR in a small simulation study based on symmetrical as well as non-symmetrical data.  Only those cases are considered which are computed in less than one hour. The computing times of the two algorithms are reported in Tables \ref{tab:timealg0norm} and \ref{tab:timealg0skewnorm}. It is seen that Algorithm~\ref{alg:median} runs slower than RR bisection  when the sample size is relatively small. But it {substantially} outperforms the RR approach as the sample size increases. In fact, when $n$ is small, $\kappa^*$ is rather easily found as its upper and lower bounds given in (\ref{eq_maxdepth}) are tight, while Algorithm~\ref{alg:median} uses much time in Step~2b to compute barycenters.
Therefore, we recommend the new Algorithm~\ref{alg:median} for larger sample sizes $n$, while the RR bisection approach is to be preferred for smaller $n$.

\begin{table}[!ht]
{\scriptsize
\begin{center}
    \caption{Average time (in seconds, over 10 runs) to compute the Tukey median with the RR bisection approach and with Algorithm~\ref{alg:median}, given $p$-variate standard normal data.}
    \label{tab:timealg0norm}
    \begin{tabular}{llrrrrr}
    \toprule
    Algorithm & $p$ & \multicolumn{5}{c}{ $n$}\\
    \cmidrule(r){3-7}
     &  & \multicolumn{1}{c}{40} & \multicolumn{1}{c}{80} & \multicolumn{1}{c}{160} & \multicolumn{1}{c}{320} & \multicolumn{1}{c}{640} \\
    \midrule
 		RR bisection & 3 & 0.031 & 0.32 & 2.35 & 47.9 & 679  \\
 		Algorithm~\ref{alg:median} & 3 & 0.034 & 0.24 & 1.93 & 41\phantom{.0} & 537 \\
 		RR bisection & 4 & 0.71\phantom{0} & 33.1\phantom{0} & 723\phantom{.00} & ---\phantom{.00} & --- \\
 		Algorithm~\ref{alg:median} & 4 & 0.87\phantom{0} & 35\phantom{.00} & 892\phantom{.00} & ---\phantom{.00} & --- \\
 		RR bisection & 5 & 16.5\phantom{00} & ---\phantom{.00} & ---\phantom{.00} & ---\phantom{.00} & --- \\
 		Algorithm~\ref{alg:median} & 5 & 25.5\phantom{00} & ---\phantom{.00} & ---\phantom{.00} & ---\phantom{.00} & --- \\
    \bottomrule
    \end{tabular}
\end{center}}
\end{table}

\begin{table}[!ht]
{\scriptsize
\begin{center}
    \caption{Average time (in seconds, over 10 runs) to compute the Tukey median with the RR bisection approach and with Algorithm~\ref{alg:median}, given $p$-variate skewed normal data as in Subsection~\ref{ssec:validation}.}
    \label{tab:timealg0skewnorm}
    \begin{tabular}{llrrrrr}
    \toprule
    Algorithm & $p$ & \multicolumn{5}{c}{ $n$}\\
    \cmidrule(r){3-7}
     &  & \multicolumn{1}{c}{40} & \multicolumn{1}{c}{80} & \multicolumn{1}{c}{160} & \multicolumn{1}{c}{320} & \multicolumn{1}{c}{640} \\
    \midrule
 		RR bisection & 3 & 0.058 & 0.99 & 17.3 & 398 & 2\,570  \\
 		Algorithm~\ref{alg:median} & 3 & 0.061 & 0.76 & 12.9 & 284 & 1\,060 \\
 		RR bisection & 4 & 2.51\phantom{0} & 347\phantom{.00} & 2\,000\phantom{.0} & --- & --- \\
 		Algorithm~\ref{alg:median} & 4 & 3.05\phantom{0} & 295\phantom{.00} & 553\phantom{.0} & --- & --- \\
 		RR bisection & 5 & 108\phantom{.000} & ---\phantom{.00} & ---\phantom{.0} & --- & --- \\
 		Algorithm~\ref{alg:median} & 5 & 142\phantom{.000} & ---\phantom{.00} & ---\phantom{.0} & --- & --- \\
    \bottomrule
    \end{tabular}
\end{center}}
\end{table}

\begin{table}[!ht]
{\scriptsize
\begin{center}
    \caption{An artificial data set for illustrating the different location of the sample mean, the coordinate-wise median and the Tukey median.}
    \label{tab:illhm}
    \begin{tabular}{lrrr}
    \toprule
    \# & \multicolumn{1}{c}{$x_1$} & \multicolumn{1}{c}{$x_2$} & \multicolumn{1}{c}{$x_3$} \\
    \midrule
1 &  1\phantom{.000} & 0\phantom{.000} & 0\phantom{.000} \\
2 & 0\phantom{.000} & 1\phantom{.000} & 0\phantom{.000} \\
3 & 0\phantom{.000} & 0\phantom{.000} & 1\phantom{.000} \\
4 & 1.5\phantom{00} & 1.5\phantom{00} & 1.5\phantom{00} \\
5 & 0.309 & 0.287 & 0.654 \\
6 & 0.733 & 0.04\phantom{0} & 0.316 \\
7 & 0.159 & 0.305 & 0.558 \\
8 & 0.056 & 0.19\phantom{0} & 0.913 \\
9 & 0.517 & 0.533 & 0.192 \\
10 & 1.012 & 0.059 & 0.099 \\
11 & 0.118 & 0.164 & 0.92\phantom{0} \\
12 & 0.175 & 0.919 & 0.222 \\
13 & 0.24\phantom{0} & 0.454 & 0.17\phantom{0} \\
14 & 0.906 & 0.056 & 0.12\phantom{0} \\
    \bottomrule
    \end{tabular}
\end{center}}
{\scriptsize
\begin{center}
    \begin{tabular}{lrrrc}
    \toprule
     location estimators & \multicolumn{1}{c}{$x_1$} & \multicolumn{1}{c}{$x_2$} & \multicolumn{1}{c}{$x_3$} & depth\\
    \midrule
       mean        &  0.480 & 0.393 & 0.476 & $\frac{1}{14}$ \\
       coordinate-wise median   &  0.275 & 0.239 & 0.269 & $0$ \\
       Tukey median &  0.454 & 0.27\phantom{0} & 0.413 &  $\frac{4}{14}$ \\
    \bottomrule
    \end{tabular}
\end{center}}
\end{table}

\begin{figure}[!hb]
\begin{center}
	\includegraphics[width=.475\textwidth,trim = 0mm 0mm 0mm 0mm,clip]{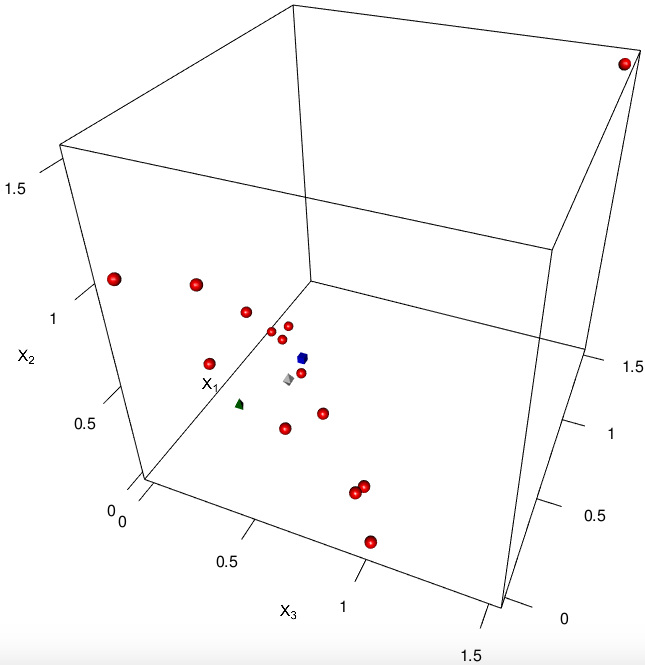}\quad\includegraphics[width=.475\textwidth,trim = 8.5mm 8.5mm 8.5mm 8.5mm,clip]{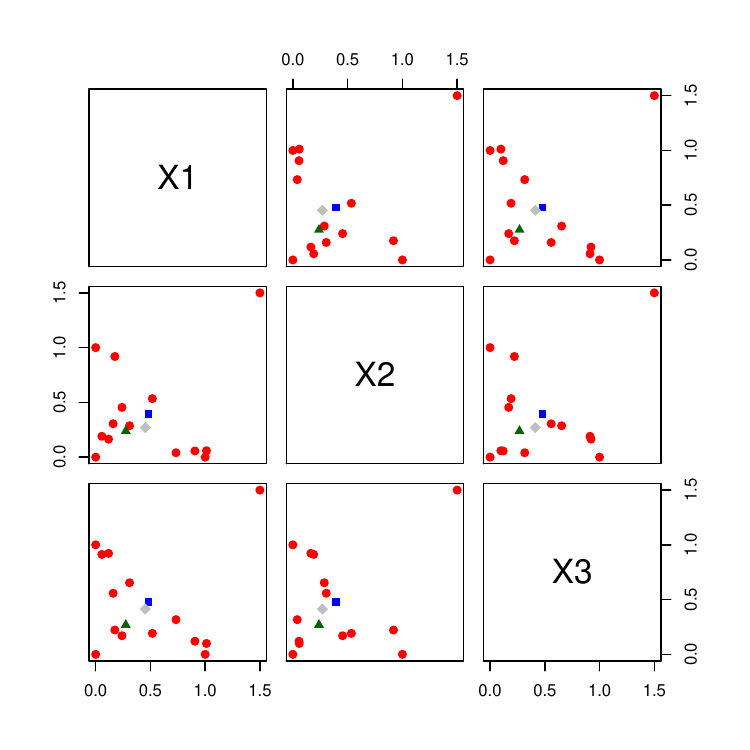}
	\caption{Data (red spheres), with mean (blue cube), coordinate-wise median (green tetrahedron) and Tukey median (gray rhombus); see Table~\ref{tab:illhm}.}
	\label{fig:samplemed}
\end{center}
\end{figure}

The positions of the sample mean, the coordinate-wise median and the Tukey median can be quite different.
To illustrate this, {we start with} a three-dimensional artificial data set in Table~\ref{tab:illhm}, which is inspired by \cite[chapter~7.1a]{RousseeuwL87}.
The data is constructed in the following way: The first three points correspond to the three canonical unit vectors. The fourth point has coordinates $(1.5, 1.5, 1.5)^\top$; it represents an outlier. Further ten points have coordinates $Z + U \cdot (\frac{1}{\sqrt{3}},\frac{1}{\sqrt{3}},\frac{1}{\sqrt{3}})^\top$, where $Z$ is a random vector having a Dirichlet distribution with parameters $(1,1,1)$ and $U$ is a random variable uniformly distributed on $[-\frac{1}{4},\frac{1}{4}]$.
The sample mean, the coordinate-wise median and the Tukey median are given in Table~\ref{tab:illhm}, together with their depths in the data cloud.
Table~\ref{tab:illhm} indicates that the coordinate-wise median lies outside the convex hull of the data set, while the sample mean -- being  sensitive to the outlier -- lies outside the convex hull of the main data (= data without {the outlier at} $(1.5, 1.5, 1.5)^\top$), which is visualized in Figure~\ref{fig:samplemed}. Moreover, the mean and the coordinate-wise median are found on different sides.
On the other hand, the Tukey median lies in the middle of the convex hull of the main data.

\begin{figure}[!ht]
\begin{center}
	\includegraphics[width=\textwidth,trim = 0mm 0mm 0mm 0mm,clip]{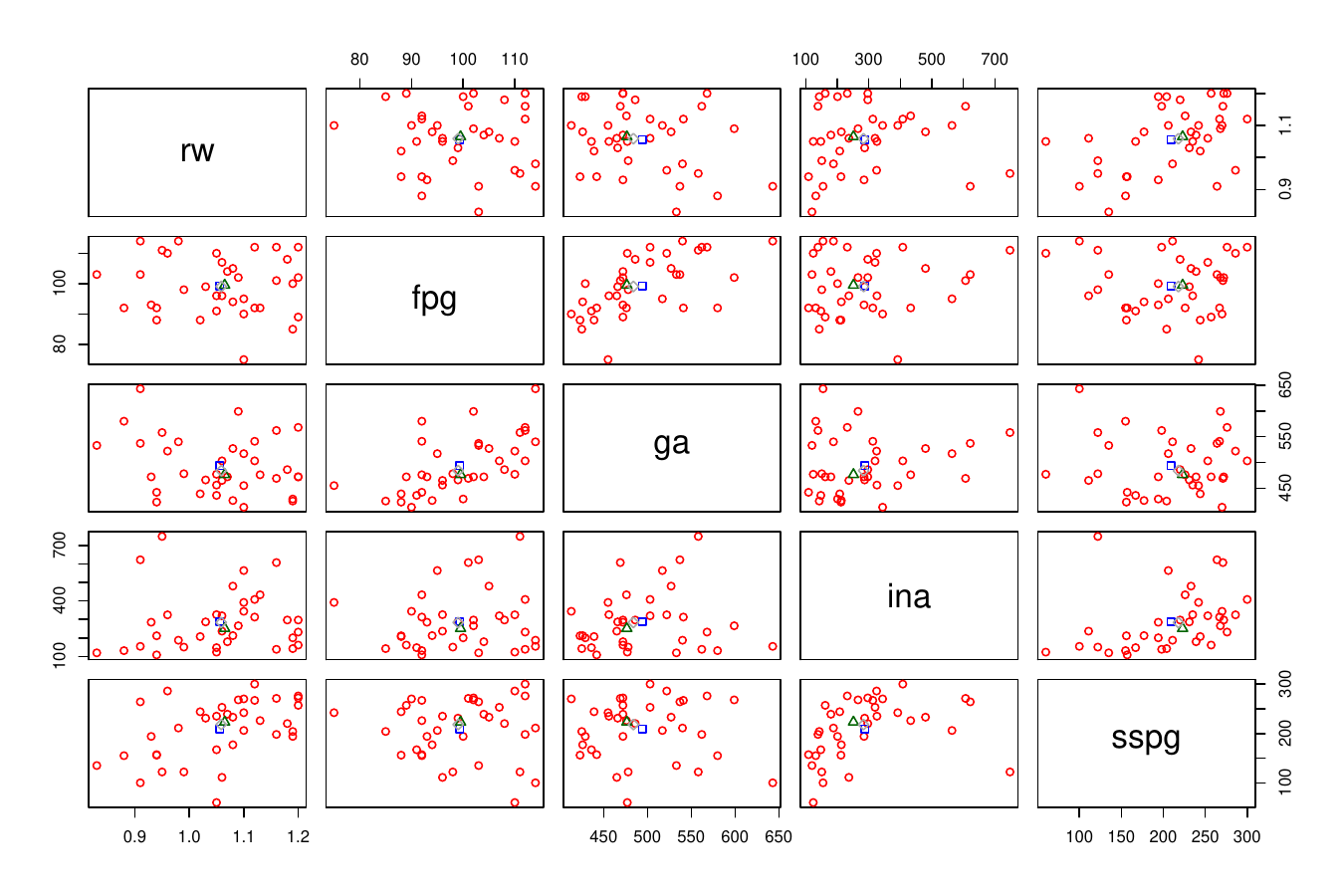}
	\caption{Data (red circles), with mean (blue rectangle), coordinate-wise median (green triangle) and Tukey median (gray rhombus) for the $36$ individuals from the Chemical Diabetes data identified as those having chemical diabetes with variables being: relative weight (rw), fasting plasma glucose (fpg), area under the plasma glucose curve (ga), area under the insulin curve (ina) and the steady state plasma glucose response (sspg).}
	\label{fig:chemdiab}
\end{center}
\end{figure}

Further, we consider the Chemical Diabetes data by~\cite{ReavenM79}. For a group of $36$ individuals having chemical diabetes, five variables are under scope: relative weight, fasting plasma glucose, area under the plasma glucose curve, area under the insulin curve, and the steady state plasma glucose response \citep[taken from \texttt{R}-package \texttt{locfit};][]{locfitRpac}. As before, we compute mean, coordinate-wise median and Tukey median for these data, and depict them in Figure~\ref{fig:chemdiab}. As we can see from the bivariate plots, again the mean (depth$=\frac{8}{36}$) and the coordinate-wise median (depth$=\frac{7}{36}$) are located away from the Tukey median (depth$=\frac{11}{36}$); the distances are $14.2$ and $33.3$, respectively.

\section{Concluding remarks}\label{Conclusions}

Two new algorithms have been constructed for computing a $\kappa$-trimmed Tukey region when $p>2$. While the first algorithm (Algorithm~\ref{alg:0}) comes without a particular search strategy, special search rules have been implemented in the second algorithm (Algorithm~\ref{alg:1}). Different from the approaches of \cite{HallinPS10} and \cite{KongM12}, Algorithm~\ref{alg:1} exploits neither cone segmentation of $\mathbb{R}^p$ nor univariate projections of the data. Employing the breadth-first ridge-by-ridge search strategy it focuses purely on the location of the data and saves both computation time on creating and searching vast structures as well as memory on storing them. Implementations employing linear programming and post-optimization (as in \cite{PaindaveineS12a,PaindaveineS12b}) occupy memory with their structures and need time for their creation and processing, which is completely avoided in Algorithm~\ref{alg:1}. Storing {a facet of a direction cone} as an average of vertices in $\mathbb{R}^p$ needs $8\, p^2$ bytes of physical memory (see \cite{PaindaveineS12a,PaindaveineS12b}),
while in Algorithm~\ref{alg:1} storing a ridge can be efficiently reduced to $\lceil\frac {p - 1}8 \log_2 n\rceil$ bytes, {which is substantially smaller} as long as
approximately (in the order of magnitude) $n < 2^{64 p}$. This explains the high computation time efficiency of Algorithm~\ref{alg:1}, which has been numerically demonstrated in Subsection~\ref{ssec:performance}. In addition, Algorithm~\ref{alg:1} finds the region's polytope and not just the relevant hyperplanes. It includes the search for an inner point, the computation of the halfspaces' intersection and the determination of the region's facets and vertices. The latter tasks can, in fact, take a major part of its computation time; see, for example, $p\ge 5$ in Table~\ref{tab:timealg1}.

Throughout the article, the data is assumed to be in general position. Violation of this assumption can be compensated by a slight perturbation of the data, which -- due to the fact that the Tukey trimmed region is defined by hyperplanes through data points --
influences location and shape of the Tukey regions just negligibly. Note that in Step~6c of Algorithm~\ref{alg:1} only regions with positive (subject to the precision constant $\epsilon$) volume are identified, otherwise they are output as empty sets.

The simulation study of Section~\ref{sec:comparisons} reveals important aspects of Algorithm~\ref{alg:1}.
Firstly, its broad numerical comparison with Algorithm~\ref{alg:0}, which is obviously correct, substantiates that Algorithm~\ref{alg:1} is correct as well. Secondly, Table~\ref{tab:timealg1} demonstrates its high speed; Table~\ref{tab:timealgs} compares its speed with that of the HPS approach. Thirdly, Table~\ref{tab:facetsalg1} and Figure~\ref{fig:ratiosfacets} show that, on an average, facets of a Tukey region are contained in much fewer hyperplanes than those found in Steps~1 to 5 of Algorithm~\ref{alg:1}. Algorithm~\ref{alg:1} shares this feature with the developments of HPS.
This still leaves room for potential improvements.

To calculate the Tukey median, which is the gravity center of the innermost Tukey region, Algorithm~\ref{alg:median} has been constructed. To our knowledge this is the first published algorithm that solves the task for dimension $p>2$.
Algorithm~\ref{alg:median} is an enriched bisection procedure that
employs Algorithm~\ref{alg:1} and an algorithm calculating the Tukey depth of a point to speed up the search for maximum Tukey depth.
Its computation time is compared with a straightforward extension of the bisection approach by \cite{RousseeuwR98}.
It turns out that the latter can be outperformed by Algorithm~\ref{alg:median} when the sample size is large.

The literature contains many other depth notions, such as the projection depth and others, which, similar to the Tukey depth, satisfy the projection property, that is, are equal to the minimum of univariate depths in any direction \citep{Dyckerhoff04}. It turns out that for several of them the trimmed regions can be computed by cutting convex polytopes with hyperplanes; see \cite{MoslerLB09} and \cite{LiuZ14b} for details. By this, similar algorithms may be constructed to calculate the respective trimmed regions in dimensions $p>2$.

Algorithms~\ref{alg:0},~\ref{alg:1} and~\ref{alg:median}, including proper visualization procedures, are implemented in the \texttt{R}-package \texttt{TukeyRegion}.

\section*{Acknowledgments}

The authors would like to thank the Editors and three anonymous Referees for their helpful comments which greatly improved the presentation of our work.

%

\end{document}